\documentclass[conference]{IEEEtran}
\IEEEoverridecommandlockouts
\usepackage{ifpdf}
\usepackage{cite}
\usepackage{comment}

\ifCLASSINFOpdf
	\usepackage[pdftex]{graphicx}
	\graphicspath{{./image/}{./image/letters/}}
\else
	\usepackage[dvips]{graphicx}
	\graphicspath{{./image/}{./image/letters/}}
\fi
\usepackage{array}
\usepackage[cmex10]{amsmath}

\newcolumntype{L}[1]{>{\raggedright\let\newline\\\arraybackslash\hspace{0pt}}m{#1}}
\newcolumntype{R}[1]{>{\raggedright\let\newline\\\arraybackslash\hspace{0pt}}m{#1}}

\usepackage{booktabs} 
\usepackage{subcaption}
\usepackage{url}
\usepackage{graphicx}
\usepackage{paralist}
\usepackage{tabularx}
\usepackage{makecell}
\usepackage{setspace}

\begin{document}

\date{}

\title{\Large \bf Screen Gleaning: A Screen Reading TEMPEST Attack on Mobile Devices Exploiting an Electromagnetic Side Channel}

\author{\IEEEauthorblockN{Zhuoran Liu, Niels Samwel, L\'eo Weissbart, Zhengyu Zhao, Dirk Lauret\IEEEauthorrefmark{1}\thanks{\IEEEauthorrefmark{1}This author is affiliated with the Eindhoven University of Technology. This work was done during an internship at Radboud University}, Lejla Batina and Martha Larson}

\IEEEauthorblockA{Institute for Computing and Information Sciences, Radboud University, The Netherlands\\ \{z.liu@cs, n.samwel@cs, l.weissbart@cs, z.zhao@cs, dirk.lauret@student, lejla@cs, m.larson@cs\}.ru.nl}}

\IEEEoverridecommandlockouts
\makeatletter\def\@IEEEpubidpullup{6.5\baselineskip}\makeatother

\maketitle

\begin{abstract}
We introduce \emph{screen gleaning}, a TEMPEST attack in which the screen of a mobile device is read without a visual line of sight, revealing sensitive information displayed on the phone screen.
The screen gleaning attack uses an antenna and a software-defined radio (SDR) to pick up the electromagnetic signal that the device sends to the screen to display, e.g., a message with a security code.
This special equipment makes it possible to recreate the signal as a gray-scale image, which we refer to as an \emph{emage}.
Here, we show that it can be used to read a security code.
The screen gleaning attack is challenging because it is often impossible for a human viewer to interpret the emage directly.
We show that this challenge can be addressed with machine learning, specifically, a deep learning classifier.
Screen gleaning will become increasingly serious as SDRs and deep learning continue to rapidly advance.
In this paper, we demonstrate the security code attack and we propose a testbed that provides a standard setup in which screen gleaning could be tested with different attacker models.
Finally, we analyze the dimensions of screen gleaning attacker models and discuss possible countermeasures with the potential to address them.
\end{abstract}
\section{Introduction}
Most of our daily business relies on the devices we carry on us.
A great deal of sensitive information is exchanged through these devices, and the security and privacy of our data is constantly at stake.
Even the task of authenticating ourselves (or our data) has  been shifted to our phones, where two-factor authentication, a common approach, requires successfully presenting two or more pieces of evidence to confirm our identity.

To protect our data, mobile devices typically use secret (cryptographic) keys that are not accessible from the outside.
Getting a hold of the key allows a hacker to steal our data.
The majority of real-world attacks on security implementations on small devices today use side-channel analysis (SCA), i.e., they measure and process physical quantities, like the power consumption or electromagnetic emanations of a chip, or reaction time of a process.
Moreover, thanks to computing power becoming ever cheaper nowadays, modern adversaries have started using state-of-the-art machine and deep learning algorithms for SCA.
Securing (embedded) systems against SCA remains a great challenge.

In certain cases, the security implementation is not the target of an attack.
Instead, the target is the sensitive information displayed on the screen.
For example, here, we can think of secret security codes sent from banks or credit card companies, giving secure access to a user who is the only one able to read the code.
SCA can take advantage of the fact that information is exposed in this way in order to mount an attack.
Since we can expect adversaries will always target the weakest link, such attacks are more feasible than cryptographic attacks i.e. cryptanalysis.

In this paper, we investigate the problem of sensitive information on mobile phone screens.
Until now, the study of side-channel analysis attacks that aim to recover the screen content of a mobile phone has focused on visible-spectrum signals.
This focus is consistent with people's general belief that protecting information on their mobile phone screen means hiding it from the line of sight of a person or a camera.
However, SCA can go beyond visible-spectrum information displayed on the screen.
In this paper, we present a low-cost SCA attack that can recover information displayed on a mobile device's screen by capturing the electromagnetic signal sent to the phone screen.
Our work introduces an attack, which we call \emph{screen gleaning}, that uses an antenna and a basic software-defined radio (SDR).
Our attack demonstrates the security threat posed by emanations leaking from mobile devices.
We release an implementation of our attacks that allows for further testing and extension.\footnote{Code available at: \url{https://github.com/cescalab/screen_gleaning}}

The side-channel analysis that we consider in this work is a type of TEMPEST technique.
TEMPEST techniques exploit vulnerabilities of communication and other types of emanations from electrical equipment that contain sensitive data~\cite{Tempest}.
From our experiments with a simple TEMPEST setup using an SDR receiver, we were able to successfully capture the phone screen content
without a visible-spectrum line of sight.
The signal recovered from the screen can be visualized as a gray-scale image, which we refer to as an \emph{emage}.
A challenge faced by our attack is that the emage is often not interpretable, meaning that it cannot be read by way of human eyesight.
We propose a machine learning-based approach capable of processing an emage that is not interpretable to the human eye
in order to recover secret information, such as a security code in two-factor authentication.

This simple attack story illustrates the potential danger of our attack:

\emph{Alice keeps her mobile phone on a stack of magazines on top of her desk. She lays the phone face down because she receives security codes and she believes that blocking the visual line of sight to the phone screen will keep the codes secure. Eve has access to Alice's desk and has hidden an antenna under the top magazine. The antenna can read the security code via electromagnetic emanations of the phone.}

In sum, this paper makes five contributions:
\begin{compactitem}
\item We present a novel side-channel technique called screen gleaning, an attack that can be used to recover information such as a security code communicated by text message. The attack does not require a visual line of sight nor the readability of the signal by a human. In fact, the signal we observe is, in most cases, not interpretable to the human eye, so the information in the leakage is not obvious.
\item We show that this kind of challenge can be tackled using machine learning, and specifically, using a deep learning classifier we are able to attain very high accuracy (of close to 90\%) when guessing the digits of a security code.
\item We quantitatively demonstrate that our attack is effective for three representative phone models under various environmental conditions. In particular, our attack is applicable in the context of cross-device, through-magazine, and noisy environments.
\item We define and validate a new testbed applicable for further research on screen gleaning. The testbed includes a parameterized attacker model, which will guide future research to systematically explore and exploit the threat of screen gleaning.
\item Finally, we propose and discuss possible countermeasures against screen gleaning attacks on mobile devices.
\end{compactitem}

The remainder of this paper is organized as follows: In Section~\ref{sec:related_work}, we discuss related work. Section~\ref{sec:attacker_model} describes the attacker model. In Section~\ref{sec:attack_setup}, we describe our measurement and machine learning setup. In Section~\ref{sec:experiments}, we explain the experiments we conducted together with the results. Section~\ref{sec:testbed} introduces a testbed.  Section~\ref{sec:discussion_countermeasures} discusses the results of the paper and describes different countermeasures.
Section~\ref{sec:text2images} discusses different formulations of the screen gleaning problem. Finally, Section~\ref{sec:conclusion} concludes the paper.

\section{Related Work} \label{sec:related_work}
\subsection{Side-Channel Attacks}
A security attack exploiting unintentional physical leakage is called a \emph{side-channel attack}. For example, an adversary might be able to monitor the power consumed by a device while it performs secret key operations~\cite{KJJ99, KJ+11}. Other sources of side-channel information, such as electromagnetic emanations from a chip~\cite{GMO01,QS01,AgrawalARR02} and timings for different operations performed~\cite{Koc96}, were also shown to be exploitable (for an overview see~\cite{MOP07}).

Side-channel attacks pose a real threat to the security of mobile and embedded devices and since their invention many countermeasures have been proposed.
The goal of countermeasures is to remove the dependence between the (secret) data and the side channel such as power consumed during the computation. An extensive study of the power side channel from mobile devices was presented in~\cite{yan2015study}.
One approach for countermeasures aims to break the link between the actual data processed by the device and the data on which the computation is performed. Such a countermeasure is usually called \emph{masking} and is exploiting the principle of secret sharing~\cite{chari-crypto99}. A second approach aims at breaking the link between the data computed by the device and the power consumed by the computations. This approach is called \emph{hiding}, and one way to achieve it is by flattening the power consumption of a device by, for example, using special logic styles that are more robust against SCA attacks such as WDDL~\cite{tiri-ches05}.

SCA attacks belong to the most serious threats to embedded crypto devices and often target the secret (cryptographic) key in a device that keeps personal data and communications secure~\cite{belgarric2016side,genkin2016ecdsa} or even white-box implementations~\cite{bos2016differential}. There are many examples of SCA attacks in the real-world such as \cite{OP11, BG+12, EKM+08} and more recent ones~\cite{MS+20, cohney2020pseudorandom, minerva}.

TEMPEST is another side-channel technique that has been known for decades. TEMPEST refers to spying on computer systems through leaking emanations, including unintentional radio or electrical signals, sounds, and vibrations~\cite{kuhn1998soft}. For example, through TEMPEST, one could easily detect a user's keystrokes using the motion sensor inside smartphones or recover the content from a computer or other screens remotely. In 1985 van Eck published the first unclassified analysis of the feasibility and security risks of emanations from computer monitors. Previously, such monitoring was believed to be a highly sophisticated attack available only to governments. However, van Eck successfully eavesdropped on a real system, at a range of hundreds of meters, by measuring electromagnetic emanations using just \$15 worth of equipment plus a CRT television set~\cite{van1985electromagnetic}. Later, Kuhn performed a comprehensive study on a range of flat-screen monitors and eavesdropping devices~\cite{kuhn2002compromising, kuhn2002optical, kuhn1998soft}. Other side channels can also convey the screen's content in the frequency range of the visible spectrum \cite{kuhn2002optical, backes2008compromising, backes2009tempest, xu2013seeing} or through acoustic channel \cite{genkin2019synesthesia} but can sometimes even require an expensive telescope.

More recently, Backes et al.~\cite{backes2008compromising, backes2009tempest} improved TEMPEST further and argue that the requirement on a direct line of sight is not necessary as they exploit reflections between the target screen and the observer. Xu et al.~\cite{xu2013seeing} broadened the scope of the attacks by relaxing the previous requirements and showing that breaches of privacy are possible even when the adversary is ``around a corner''. A new technique is presented for reconstructing the text typed on a mobile device, including password recovery via analysis of finger motions over the keyboard and language model. The main distinction from the works by Backes et al. is that they use ``repeated'' reflections, i.e., reflections of reflections in nearby objects, but always originating from the surface of a person's eyeball. Nevertheless, all those papers use direct or indirect reflections from the screen, which makes their research line very different from ours. More specifically,  those papers focus on recovering text and images from the screen while being typed and being captured by a camera from an eyeball, which implies rather special assumptions on the setup and attacker model.

Hayashi et al. performed a comprehensive evaluation of electromagnetic emanations from a chip including countermeasures~\cite{hayashi2012efficient,hayashi2012analysis,hayashi2014threat,hayashi2016remote}. However, their focus is on recovering secret information from ``inside'' such as cryptographic keys and not the screen content.

As a follow-up, the work of Kinugawa et al.~\cite{kinugawa2019electromagnetic} demonstrates that it is possible to amplify the electromagnetic leakage with cheap hardware modification added on potentially any device and spread the attack to a broader distance. They demonstrate that this additional circuitry, a so-called interceptor, enlarges the amount of leakage and even forces leakage in devices that do not suffer potential electromagnetic leakage.

Goller and Sigl proposed to use standard radio equipment when performing side-channel attacks on smartphones~\cite{goller2015side}. They also aimed their attack at cryptographic operations inside the chip as they demonstrate the ability to distinguish between squaring and multiplications. This observation could lead to the full RSA key recovery, assuming that the modular exponentiation is implemented with a basic square-and-multiply algorithm. Their setup used an Android phone to collect electromagnetic leakages from (albeit they modified the hardware, which makes their attacker's model different).

There exist many papers considering finger movements on the screen or other traces from typing on a smartphone. For example, Cai et al. developed an Android application called TouchLogger, which extracts features from device orientation data to infer keystrokes~\cite{cai2011touchlogger}.
Aviv et al. used the embedded accelerometer sensor to learn user tapping and gesturing to unlock smartphones~\cite{aviv2012practicality}. In another work, they introduce smudge attacks as a method that relies on detecting the oily smudges left behind by the user's fingers when operating the device using simple cameras and image processing software~\cite{aviv2010smudge}.

As another two examples of recent work, we also mention the papers of Genkin et al.~\cite{genkin2015get, genkin2019synesthesia}.
In~\cite{genkin2015get}, the authors use various side channels like power and electromagnetic radiation to extract cryptographic keys, i.e., RSA and ElGamal keys from laptops, but do not discuss the possibility to perform the attacks on a phone.
On the other hand, in~\cite{ genkin2019synesthesia} the authors show how to extract the screen content by using the acoustic side channel. They demonstrate how the sound can be picked up by microphones from webcams or screens and transmitted during a video conference call or archived recordings. It can also be recorded by a smartphone or other device with a microphone placed in the screen proximity.
These two examples are different from our work because they use either another kind of emanation or have different attack goals (or both).

Other work using acoustic side channels is from Berger et. al~\cite{berger2006dictionary}, which demonstrated a dictionary attack using keyboard acoustic emanations. Backes et al.~\cite{backes2010acoustic} investigated acoustic side channel on printers, and Asonov and Agrawal~\cite{asonov2004keyboard} used the sound emanated by different keys to recover information typed on a keyboard.

In sum, the uniqueness of our contribution is a side-channel analysis attack that exploits the electromagnetic emanations of the display cable from a mobile phone.
These emanations are less accessible and may be substantially weaker than the signals analyzed in more traditional TEMPEST technique attacks.
To the best of our knowledge, the most recent work, which bears superficially similarity to ours, is~\cite{lemarchand2020electro}.
This work applied deep learning to recognition on TEMPEST signals, but does so with the goal of automation and enhancement.
In other words,~\cite{lemarchand2020electro} targeted a captured signal in which the content is clearly interpretable to the human eye (cf. Figure~2 in~\cite{lemarchand2020electro}).
In our work, machine learning is used for the purpose of identification.
We face the challenge of an uninterpretable emage derived from a mobile phone.

\subsection{Deep Learning and Side-channel Analysis}
Several side-channel analysis techniques are based on profiling a physical device and are commonly known as \emph{template attacks} and refer to the first such attack presented by Chari et.al.~\cite{chari-ches2002}. Profiling attacks estimate a power profile of a cryptographic device for each possible secret key from their resulting power traces (also known as the training phase) and predicting the corresponding key of an unknown trace. From this very similar approach to machine learning, several methods have been inspired by machine learning and neural networks~\cite{carbone2019deep,kim2019make,CDP17,MPP16}. These methods have raised much attention as they provide more powerful attacks than the state-of-the-art.
In our work, we will discuss the usability of deep learning, specifically Convolutional Neural Networks (CNNs)~\cite{lecun1998gradient,krizhevsky2012imagenet}, for classifying the emages that are reconstructed from the screen content.

Image classification is the task of predicting a class for a given image according to its content.
In the context of machine learning, it can be automated by modeling a transformation from an image to its corresponding class.
Early research~\cite{van2009evaluating,grigorescu2002comparison,lowe2004distinctive} tackled this problem via a two-step process: manually extracting features from the images and then training a discriminative model for classification.

Deep learning algorithms, such as CNNs~\cite{lecun1998gradient,krizhevsky2012imagenet}, automatically learn image features simultaneously with learning the classification by making use of a large number of filters in an end-to-end manner.
Deep learning has lead to breakthrough success in general image classification.
Large-scale training and diverse data augmentation techniques make an important contribution.
In particular, it has been demonstrated that deep learning can achieve superhuman performance in specific domains where the discriminative visual patterns are hard to distinguish by the human eye, e.g., image forensics and steganalysis~\cite{ye2017deep,bayar2018constrained,zhou2018learning}.
In our work, since the emage content is hardly recognizable to the human eye, we use CNNs to capture the subtle differences between various classes, rather than relying on human-interpretable features.


\section{Attacker Model} \label{sec:attacker_model}
\begin{figure}[t]
\centering
\includegraphics[width=\linewidth]{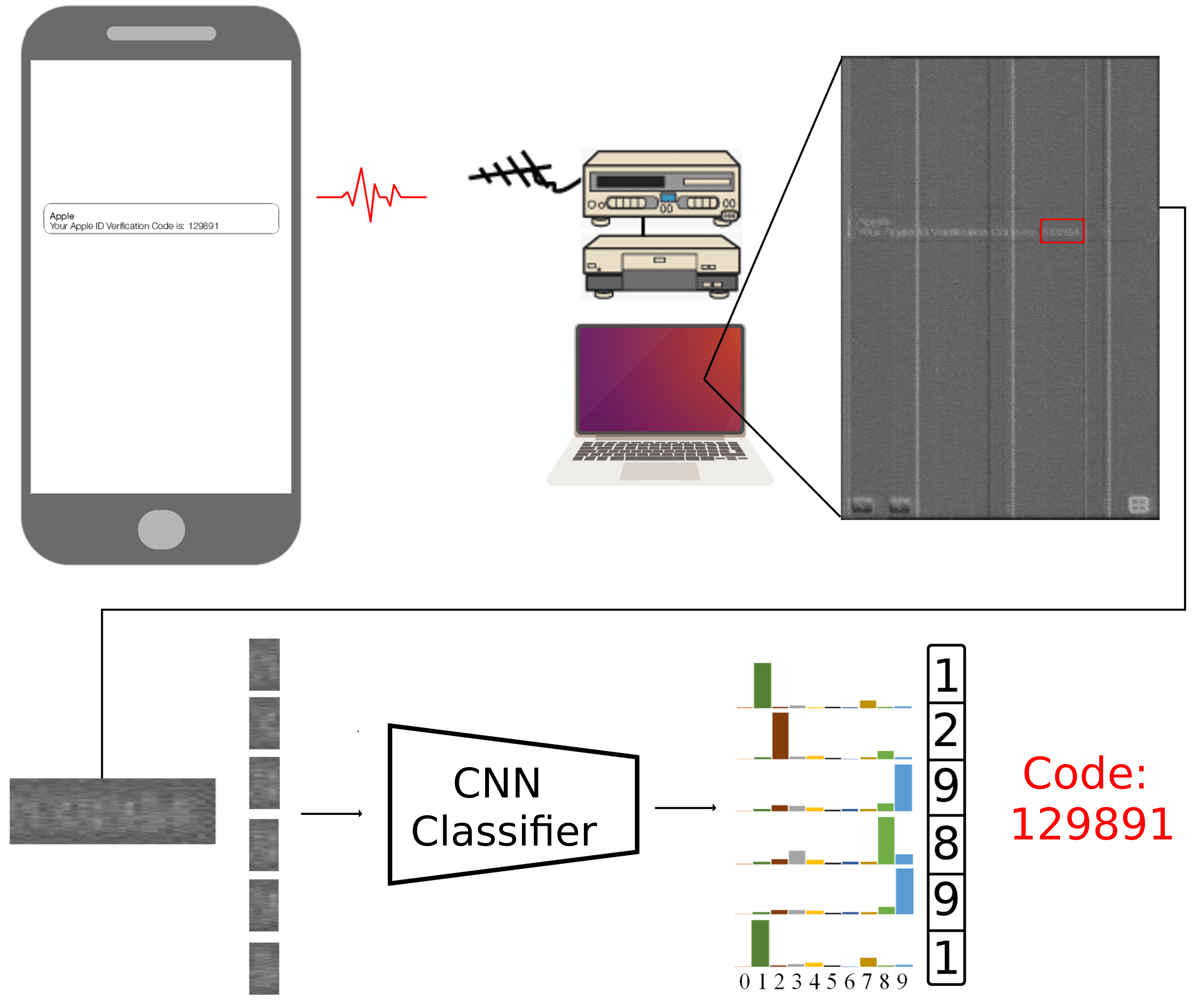}
\caption{Screen gleaning attack. The target emits electromagnetic side-band intercepted by an antenna connected to a software-defined radio (SDR). The leaked information is collected and reconstructed as a gray-scale image (emage). From emage, the 6-digit security code is cropped and fed into a CNN classifier for recognition.\label{fig:attackmodel}}
\end{figure}

The attacker's goal is to recover the information (e.g., security code, password, or message) displayed on the target display.
We start from the general attack story presented in the introduction: an antenna is planted that can read a security code from a mobile phone screen without a visible-spectrum line of sight.
This story is the basis for the attacker model, which is illustrated in Figure~\ref{fig:attackmodel} and characterized in detail in Table~\ref{tbl:our_attacker_model}.
In this section, we provide an explanation of the attacker model and its motivation.

Our attacker model makes the following assumptions:
\begin{compactitem}
	\item The set of symbols displayed on the phone is finite and known (i.e., digits 0-9). This assumption holds true of any information expressed as alphanumeric characters.
	\item The attacker has access to a profiling device sufficiently similar to the target device, which is used to collect training data for the machine learning classifier.
	\item The context for the attack is a side-channel analysis setup for a passive adversary, featuring an antenna that has been positioned to collect electromagnetic emanations and an SDR device for signal processing. The antenna picks up the signal from close range.
	\item During the attack, the attacker can collect electromagnetic traces from the target device representing the image displayed on the screen. The traces are analyzed for the appearance and identification of the pincode.
\end{compactitem}

We now explain the attack in more detail.
The device under attack (Figure~\ref{fig:attackmodel} upper left) is assumed to be a standard device (e.g. a phone) and comply with the standards imposed by EMC regulations laws.
The attacker can only rely on unintentional electromagnetic leakage of the device under attack to reconstruct the image displayed on the victim's screen.
The leaked electromagnetic signal is characterized by several physical properties of the screen (e.g., resolution, refresh rate) and by the technology used for the image rendering (e.g., CRT, TFT-LCD).
The work of Marinov~\cite{marinov14} led to the development of a software toolkit (Figure~\ref{fig:attackmodel} upper middle) capable of reconstructing the image from emanations of a video monitor. This tool, TempestSDR, is publicly available~\cite{marinovgit} and used as a starting block of our work.

It is important to understand that the challenges involved with the capture and interpretation of electromagnetic emanations from the display cable of a mobile phone are different from those with other devices considered in conventional TEMPEST studies.
Given the advance in video display technology, modern screens now use less energy and their circuitry is getting smaller.
The resulting electromagnetic coupling is lowered and the carrying frequency of the electromagnetic emanation is increased.
Additionally, basic design compliance to guarantee the electromagnetic compatibility of the products helps to reduce unintentional leakages.
These factors make the exploitation of this signal more complex and degrade the intercepted signal of electromagnetic emanation of cables.

For completeness, we discuss the future implications of the choices made in our setup.
Here, we choose to work in close range and use a near-field magnetic probe.
We note that in the future, additional effort can be invested in order to design the antenna that takes into account the electromagnetic properties of the leaking device.
A broad description of these characteristics and how to select a matching antenna to the electromagnetic leakage is discussed in~\cite{kuhn2002compromising}.
We assume that better antennas will relax the constraints of our attacker model in the future.
Some relevant work about designing antennas for a better electromagnetic setup was done in~\cite{sekiguchi2013study}.

\begin{table}[t]
\normalsize
\newcommand{\tabincell}[2]{\begin{tabular}{@{}#1@{}}#2\end{tabular}}
\resizebox{\columnwidth}{!}{
\begin{tabular}{L{.23\linewidth} R{.77\linewidth}}
\toprule
\textbf{Dimension}&\textbf{Description} \\\midrule
Message&A six digit security code; each content digit can be 0-9 with equal probability.\\\midrule
\tabincell{l}{Message\\appearance} & The standard size, position, and font with which a security code appears as a push message during a conventional authentication procedure. Plain background and standard brightness are used. \\
    	\midrule
    	\tabincell{l}{Attack\\hardware} & Close field antenna and standard SDR; we assume immediate proximity of the antenna. \\
    	\midrule
    	\tabincell{l}{Device\\profiling} & We assume full access to the profiling device for the purpose of collecting training data; We can display an image on the device. We have sufficient time to collect data from several sessions. (2-3 hours.) \\

    	\midrule
    	\tabincell{l}{Computational\\resources}& About 24 hours on a standard laptop, or 1 hour on a laptop with a GPU for training. For recovery, once the emage has been captured, a matter of seconds. \\
		\bottomrule
	\end{tabular}
	}

	\caption{Five-dimensional attacker model: Specifications of the attacker model used in our security code attack}
		\label{tbl:our_attacker_model}
\end{table}
We next turn to discuss the ``profiling stage'' of the attack in more detail.
As previously mentioned, if an emage has a low signal-to-noise ratio (SNR), it is impossible for the attacker to read the emage with the naked eye.
In this case, in order to interpret the image and recover the screen content, the attacker must use machine learning to analyze and interpret the emage.
To realize the machine learning classifier, it is necessary to train it on examples of the signal of the antenna, which is the ``profiling'' part of our attack.
The attacker uses the profiling device to display specific images with known content and captures the emages that correspond to these emages.
The collected emages are labeled with the image content and constitutes the training data set.

Once the model is trained, the attacker will be able to record emages from the device under attack to derive the secret information displayed.
The process is illustrated in Figure~\ref{fig:attackmodel}.
The success of the attack is measured as the classification accuracy, which quantifies the ability of the classifier to recover the six-digit security code.

In our experiments, we first set up our attack using the same device at the profiling device and the target device.
Considering the same target for profiling and attack phases allows us to understand the danger of the attack under best-case conditions for training data collection.
Later, we extend the attack to using two different devices.
We consider a device of the same make and model to collect data, and also the situation in which the profiling device is another phone altogether.

We close this section by explicitly summarizing the difference between our attack model and those previously studied in the literature.
Because of the specific challenges of mobile phones discussed above, the types of attacks that are successful are not the same as the attacks previously discussed in the literature for other devices.
While the TEMPEST technique has been known for decades, there have been no demonstrations of it on mobile devices.
The attack model for mobile phones until now has assumed the exploitation of reflections of a visible-spectrum signal, which means that the information is supposed to be visually accessible to humans~\cite{backes2008compromising, backes2009tempest}.
Other attack setups exploiting electromagnetic side-band have the goal to do key recovery from cryptographic implementations running on the phone~\cite{goller2015side}.
Our work is different as it shows for the first time the threat of TEMPEST on a range of mobile phones for a (machine learning-assisted) adversary that can extract the screen content that could appear incomprehensible to humans.

In Section~\ref{sec:testbed}, we will provide additional discussion of the attacker model, describing how future work can build on and extend it.
We emphasize that the attack that we present in this section is important because it reveals the danger in anticipation of the development of more sophisticated attackers.

\begin{figure}[t]
    \centering
    \includegraphics[width=\linewidth]{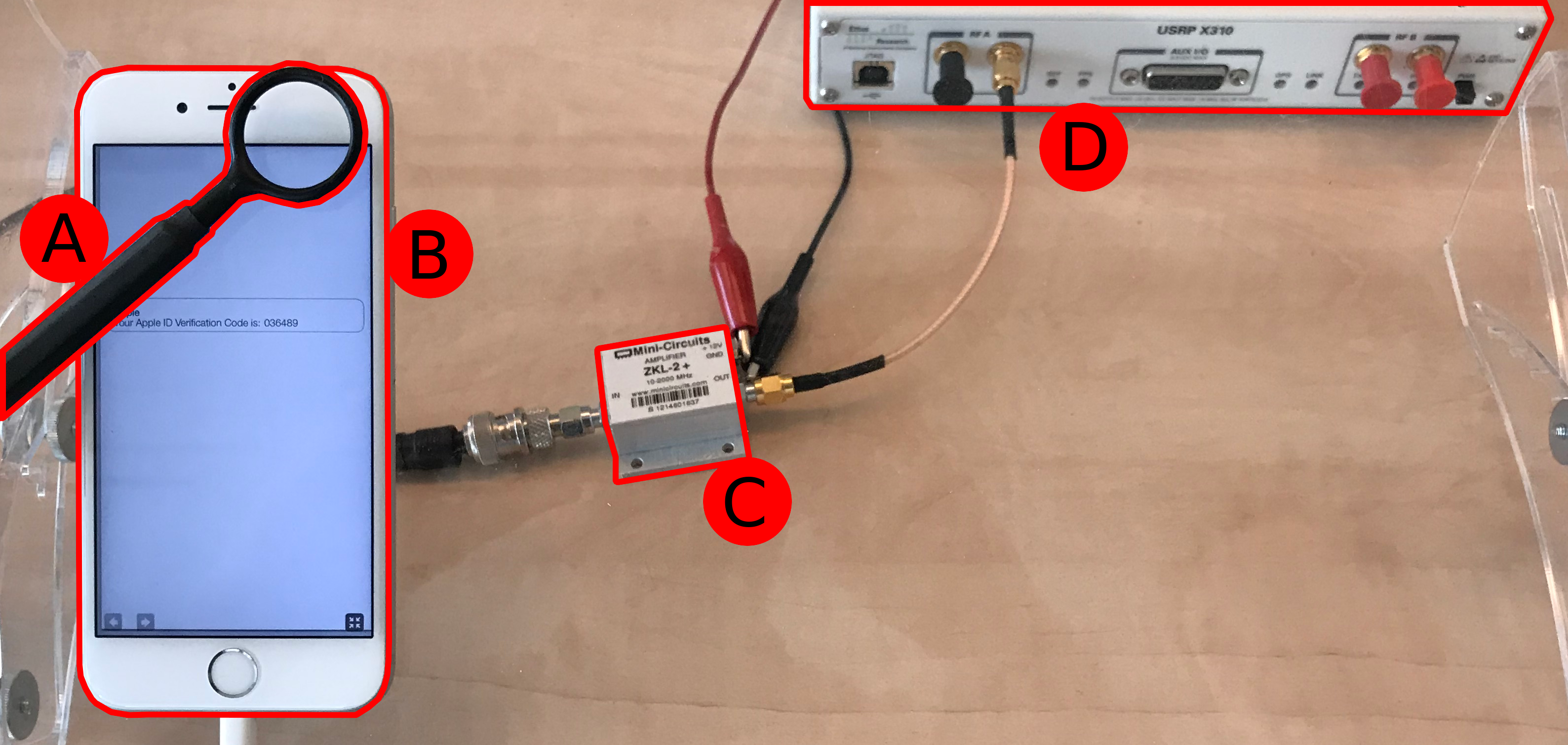}
    \caption{Measurement setup. (A) near-field probe, (B) targeted phone displaying a security code, (C) power amplifier, and (D) the software-defined radio.}
    \label{fig:setup}
\end{figure}

\section{Attack Setup} \label{sec:attack_setup}

\subsection{Measurement Setup}

\subsubsection{Target}
A TEMPEST attack can potentially be performed on any communication device, whether mechanical or electrical, as long as the signal involved for the communication can be intercepted by a third party using unconventional means. It is non-trivial to define such a means, and also the cause of the communication leakage, because this leak has not been designed. Leaks have been shown in the literature to be of several forms linked to the physically inherent properties of the communication signal.

Our work focuses on electronic personal mobile devices leaking an analogue video signal as electromagnetic emanation.
The signal leaks from the ribbon cable that connects the graphical computing unit to the screen.
Note that the attack we studied here would be blocked in the case that video encoding is applied to the video signal.
The vulnerability of encoded signals needs to be investigated in future work.

The cable, which conveys the electric information, acts as an undesired antenna and transmits the video signal in the electromagnetic spectrum in the surrounding area.
An impedance mismatch between the cable and socket on both the motherboard and the display can enhance the ribbon cable's leakage. The difference of impedance is possibly caused by a dimension mismatch between the socket and the ribbon cable. The connecting cable is often designed to be smaller than the socket to avoid possible interference between neighboring connectors. Since each manufacturer is free to use a different offset for these cables, different phones radiate with varying signal strengths. Future research should prove the hypothesis that different phones have different signal strengths radiated, by means of quantifying the radiated signal.
According to~\cite{marinov14}, the frequency of the leaked signal is dependent on several screen properties and can be estimated at a specific frequency (and its harmonics) with the following relation: $f_v=x_t \times y_t \times f_r$, where $x_t$ and $y_t$ are respectively height and with of the screen in pixels and $f_r$ is the screen refresh rate in Hertz (Hz).

\begin{table*}[t]
\newcommand{\tabincell}[2]{\begin{tabular}{@{}#1@{}}#2\end{tabular}}
    \centering
    \begin{tabular}{l|cccc}
\toprule
\textbf{Phone}&\textbf{Leakage Center Frequency}&\textbf{SNR}&\textbf{Screen Technology}&\textbf{OS}\\\midrule
iPhone~6s&295~MHz&33.4dB&IPS LCD&IOS 10.2.1\\
iPhone~6-A&105~MHz&25.0dB&IPS LCD&IOS 12.4.8\\
iPhone~6-B&105~MHz&26.8dB&IPS LCD&IOS 12.4.8\\
iPhone~6-C&105~MHz&24.9dB&IPS LCD&IOS 12.4.8\\
Honor~6X&465~MHz&36.6dB&IPS LCD&Android 7.0\\
Samsung Galaxy A3&295~MHz&25.9dB&AMOLED&Android 5.0\\
\bottomrule
    \end{tabular}
    \caption{Screen specification of the targets}
    \label{tab:target_specs}
\end{table*}

The principal target in the experiment section is an Apple iPhone~6s with an IPS LCD screen of size 1334 $\times$ 750 pixels. We also present results using different targets to prove the portability of the attack. The different targets used are listed in Table~\ref{tab:target_specs} with the center frequency of the strongest video signal leakage, the SNR of the leakage as well as relevant information about the targets (screen size, technology and Operating System version).
The SNR is computed at the center frequency of the signal with a bandwidth of 50~MHz and a resolution of 25~kHz.

\subsubsection{Equipment}

Figure~\ref{fig:setup} shows an overview of the setup with the elements labeled as follows.
The antenna we use is a passive Langer RF-R 400 magnetic probe (A).
The target is an iPhone~6s (B).
The signal from the probe is amplified with a Minicircuits ZKL-2 amplifier (C) and digitized with a Software-Defined Radio (SDR), an Ettus X310 (D) with a UBX-160 daughter-board.
The signal acquired by the SDR is then interpreted with TempestSDR~\cite{marinovgit}, an open-source tool capable of reconstructing an image from the display by the obtained sequence of electromagnetic leakages~\cite{marinov14}.

\subsubsection{Positioning and Parameters}
We use SCA equipment to show a proof of concept of this attack because the parameters and positioning settings are close in the two contexts.
Nonetheless, using more specialized equipment for TEMPEST attacks may achieve better results.
The magnetic probe is placed on top of the target, at a close distance ($<1$cm).
The best position and distance of the probe from the target is manually optimized to observe the best possible signal to noise ratio (SNR).

TempestSDR has a number of parameters to configure the SDR and to recover the image from the signal.
The SDR has the following parameters: center frequency, bandwidth, and sampling rate.
The bandwidth and sampling rate are fixed to 12.5~MHz and 25 M samples per second respectively.
The SDR captures a bandwidth of 12.5~MHz around the adjustable center frequency.
We adjust the center frequency to determine the best SNR.
The parameters to recover an image from a signal are: height and width in pixels and refresh rate in frames per second.
There are also sliders to adjust the gain and low pass filter of the SDR.
The values for the width and the height do not necessarily correspond to the dimensions of the screen as more pixels may be transmitted than those that are displayed.
The selected refresh rate should be the closest possible to the actual refresh rate and can be configured with high precision in the software.
The parameters require high precision and differ among devices, they should be determined following the description in~\cite[Section 4.2]{marinov14}.

\subsubsection{Automation}
\begin{figure}[t]
\centering

\includegraphics[width=\linewidth]{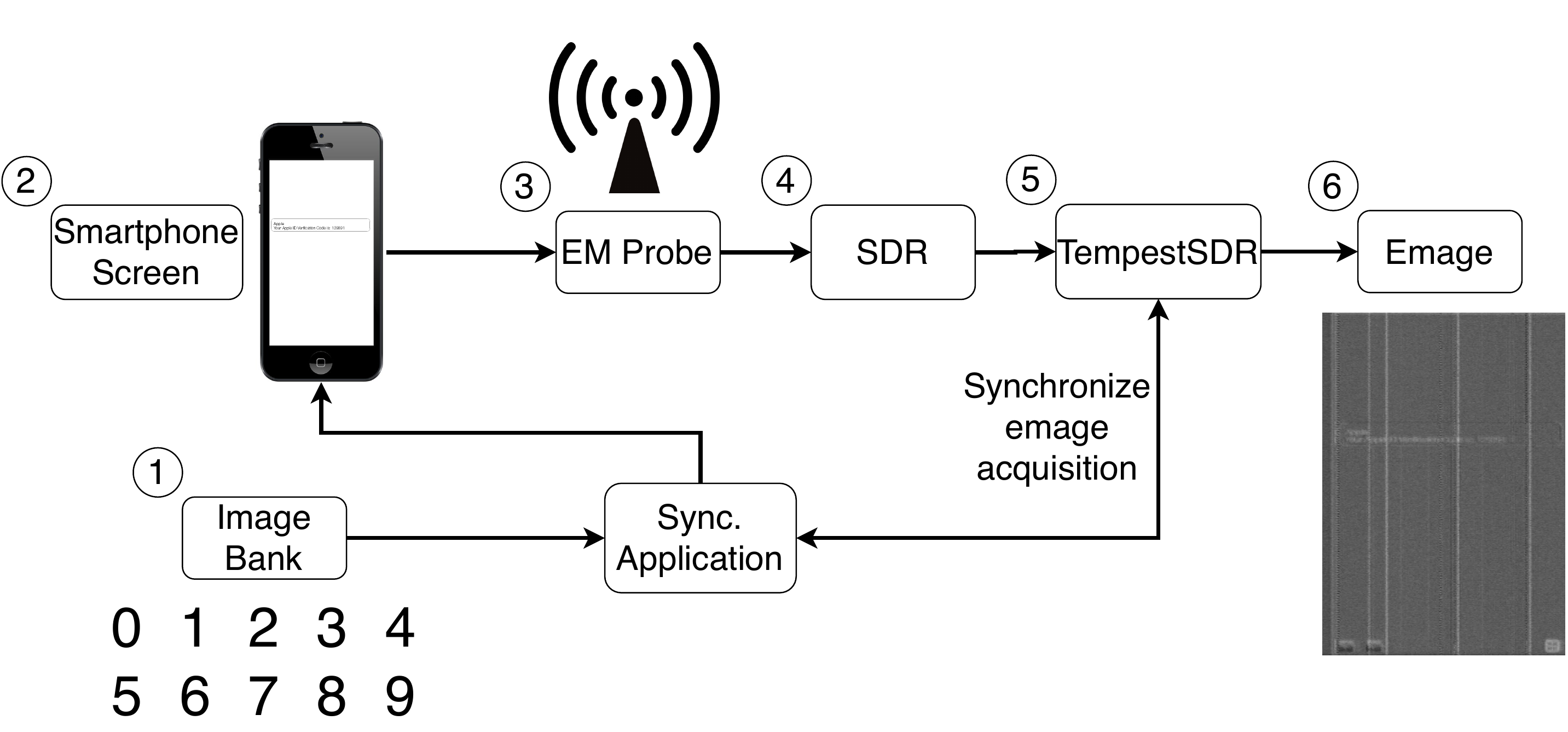}
\caption{Automation workflow\label{fig:automation}}
\end{figure}

The TempestSDR software contains a built-in function to store a processed frame. The image captured from the reconstruction of the frame is called the emage.
For timing efficiency and reliability of the capturing process, we use an automated approach to emage acquisition.
Specifically, we set up an application that synchronizes the selection of an image in the image bank, displays it on the screen and saves the emage (see Figure~\ref{fig:automation}).
This application consists of a Javascript server and a simple website.
Additionally, a small modification to the TempestSDR software was made to automatically save images and communicate with the server.
The TempestSDR sends a signal to the server to display an image from the image bank.
The server communicates this to the webpage loaded on the phone and the webpage reports back when the image is changed.
The TempestSDR captures a parametrizable number of emages and asks for a new image.

\subsection{Machine Learning Setup}
\label{subsec:ML-setup}
Here, we describe the collecting process of emage data sets used to train our security code classifier.
Given an emage from the device under attack, the classifier can produce a prediction of the message, which contains a six-digit security code, displayed on the smartphone screen.

It is important to note that the attacks we investigate here can be formulated within a discrimination scenario.
This means that the goal of the attack is to discriminate between a set of messages about which the attacker has full information.
For example, in the security code scenario, the attacker knows that the security code consists of six places and the symbol in each of those places is a digit from 0-9.
It is important to contrast the discrimination scenario with a \emph{reconstruction scenario}.
The scenarios differ in the amount of information about the content of the screen available to the attacker.
It is also possible to formulate screen gleaning attacks within a reconstruction scenario.
Here, the goal is to recover the content of the screen exactly as displayed on the screen without using any prior knowledge of what content might be displayed.
The reconstruction problem will be discussed further as an outlook onto future work in Section~\ref{sec:text2images}.

\subsubsection{Data Collection}
\begin{figure}[t]
    \centering
    \includegraphics[width=0.7\columnwidth]{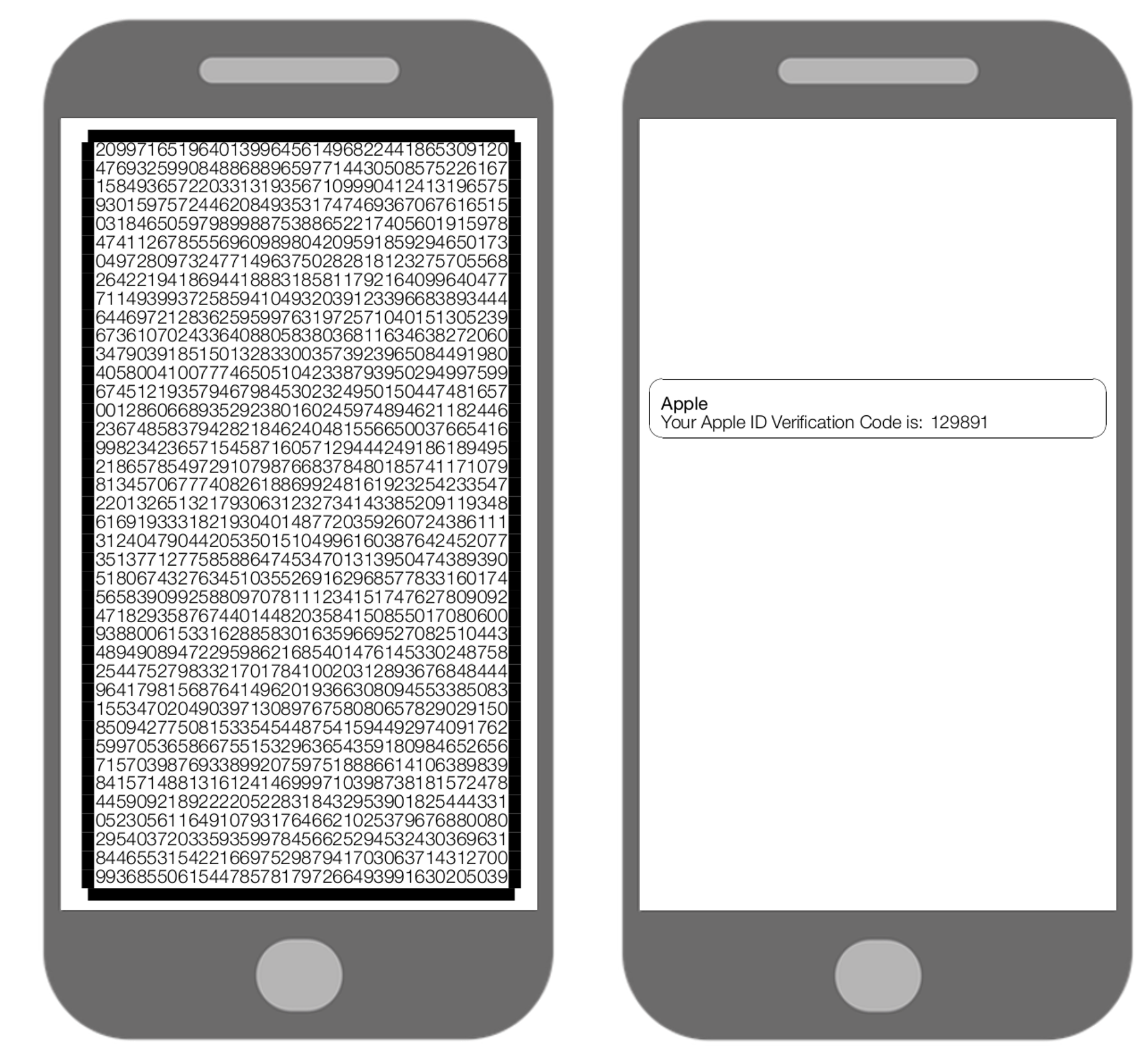}
    \caption{Screen display used to collect digits from a multi-crop grid for training our classifier (left) and from an automated text message containing a security code for testing (right).\label{fig:screen_display}}
\end{figure}
To train a classifier, the attacker needs to collect training data from the same distribution as the practical data shown on the target device or from similar types of data from other devices.
Practically, collecting security code data directly from text messages needs a large amount of annotation effort, since people have to inspect each message and crop the code one by one.
Considering such inconvenience, we propose to generate images depicting different numbers (0-9) over the whole image, and collect data using a multi-crop approach.
Specifically, each single image is split into $40\times40=1600$ cells of digits, as shown in Figure~\ref{fig:screen_display} (left).
Accordingly, after each trial of emage generation, we can get 1600 emages of different digits.
We crop the instances with a certain human inspection to guarantee the data quality.
We conduct multiple sessions of emage generation to alleviate the influence of distribution shift, which is validated effective as in Section~\ref{subsec:pin_train}.

\begin{table*}[t]
    \centering
    \begin{tabular}{ll|ccccc}
    \toprule
        \textbf{Data}&\textbf{Displayed Content}& \textbf{iPhone~6s} & \textbf{iPhone~6-A} & \textbf{iPhone~6-B} & \textbf{Honor~6X} \\\midrule
        Train/Val/Test&Multi-Crop Grid Data& 10 & 5 & N/A & 5 \\
        Single-device Test&Security Code Data& 2 & 2 & N/A & 2 \\
        Cross-device Test&Security Code Data& N/A & N/A & 1 & N/A \\
        Magazine Test&Security Code Data& 1 & 1 & N/A & 1 \\
        Noise Test&Security Code Data& 1 & 1 & N/A & 1 \\
    \bottomrule
    \end{tabular}
    \caption{The list of collected data sessions for different phones in the security code attack. Multi-crop grid data represents the data collected in the case of multi-crop, and security code data represents the simulated text message with the security code.}
    \label{tab:phonedata}
\end{table*}

\begin{table*}[t]
    \begin{singlespace}
    \renewcommand{\arraystretch}{1}
    \centering
    \begin{tabular}{l|ccccccccccc}
        \toprule
        \textbf{Digits}&0&1&2&3&4&5&6&7&8&9&All\\
        \midrule
       \textbf{Acc. (\%)}&87.2&86.8&97.4&75.8&99.1&97.4&95.1&93.1&82.5&86.1&89.8\\
        \bottomrule
    \end{tabular}
    \end{singlespace}
     \caption{Accuracy with respect to different digits (0-9) and overall accuracy in our security code attack.}
    \label{tab:pin_diff_overall}
\end{table*}

\begin{figure}[t]
	\centering
	\includegraphics[width=0.8\linewidth]{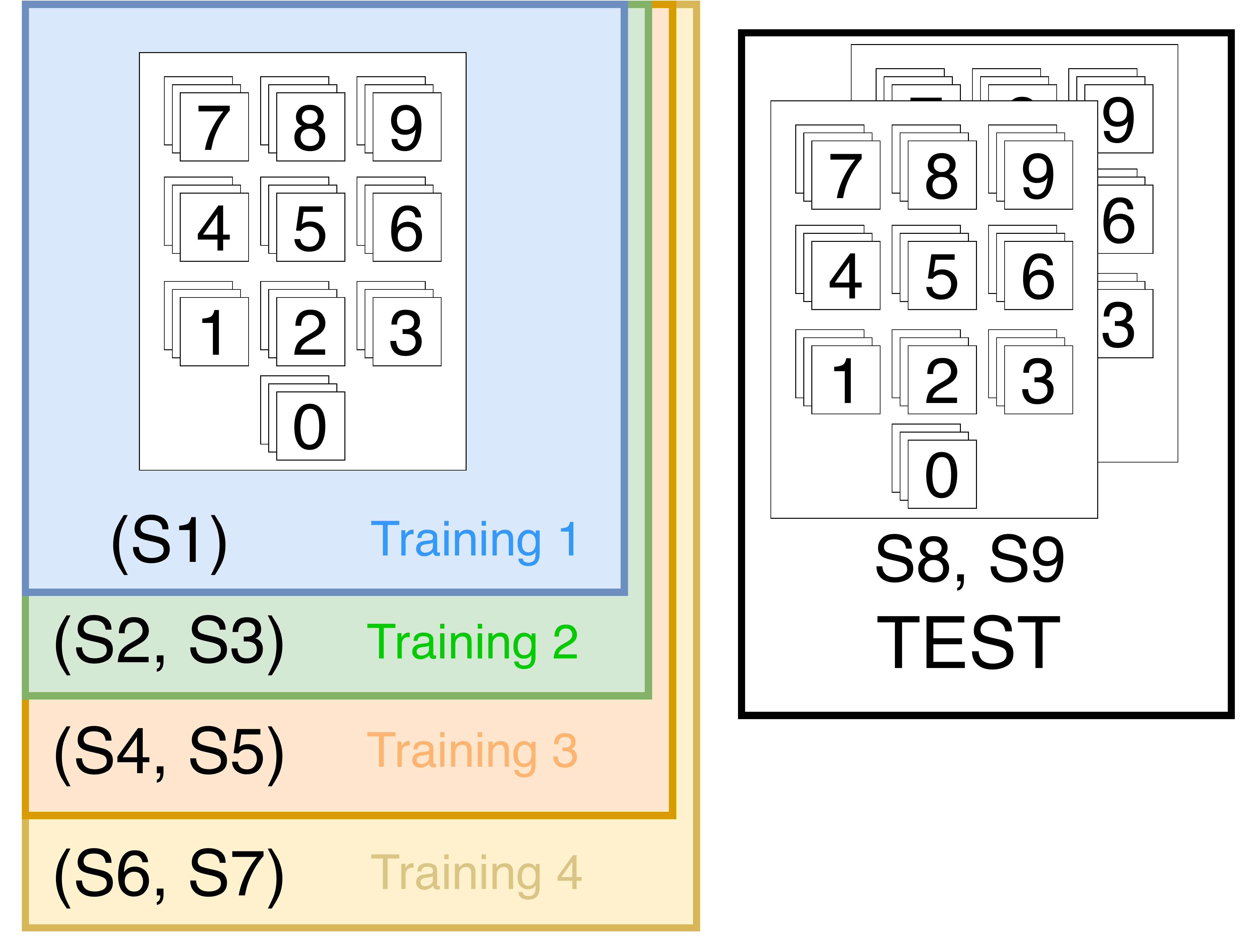}
	\caption{Train/test splits specified in the case of multi-crop grid, where the training set is gradually enlarged by including more sessions, and the test set is fixed with two sessions.\label{fig:sec_batch}}
\end{figure}

\subsubsection{CNN Architecture and Model Training}
\begin{figure}[t]
	\centering

	\includegraphics[width=\columnwidth]{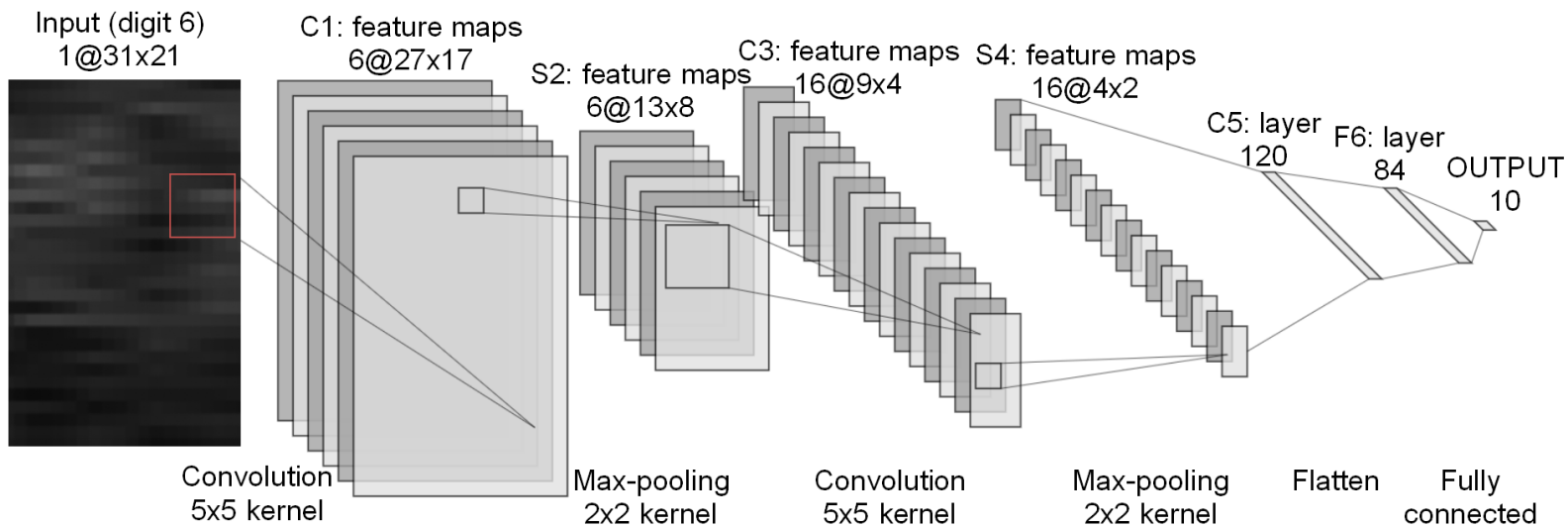}
	\caption{CNN architecture used in our security code attack.\label{fig:CNN_structure_pincode}}
\end{figure}

For the model architecture, we adopt the simple LeNet~\cite{lecun1998gradient}, which was initially proposed for handwritten digit recognition. We slightly adapt the LeNet to fit our input emage size of $31\times21$ (for Honor 6X the input size is $45\times21$, and for iPhone 6 is $31\times20$).
The details of the architecture is shown in Figure~\ref{fig:CNN_structure_pincode}.
The PyTorch is used for our implementation and the experiments are run on a workstation with a 16-core CPU and a GTX1080Ti GPU.
In all cases, 80\% of multi-crop grid data are used for training, 10\% for validation and 10\% for testing.
Each round of training can be finished within one hour when using the Adam optimizer~\cite{kingma2014adam} with a learning rate of 0.001.
We conduct the training over 100 epochs with a batch size of 256, and select the optimal model based on the validation accuracy.

\begin{figure}[t]
	\centering

	\includegraphics[width=\columnwidth]{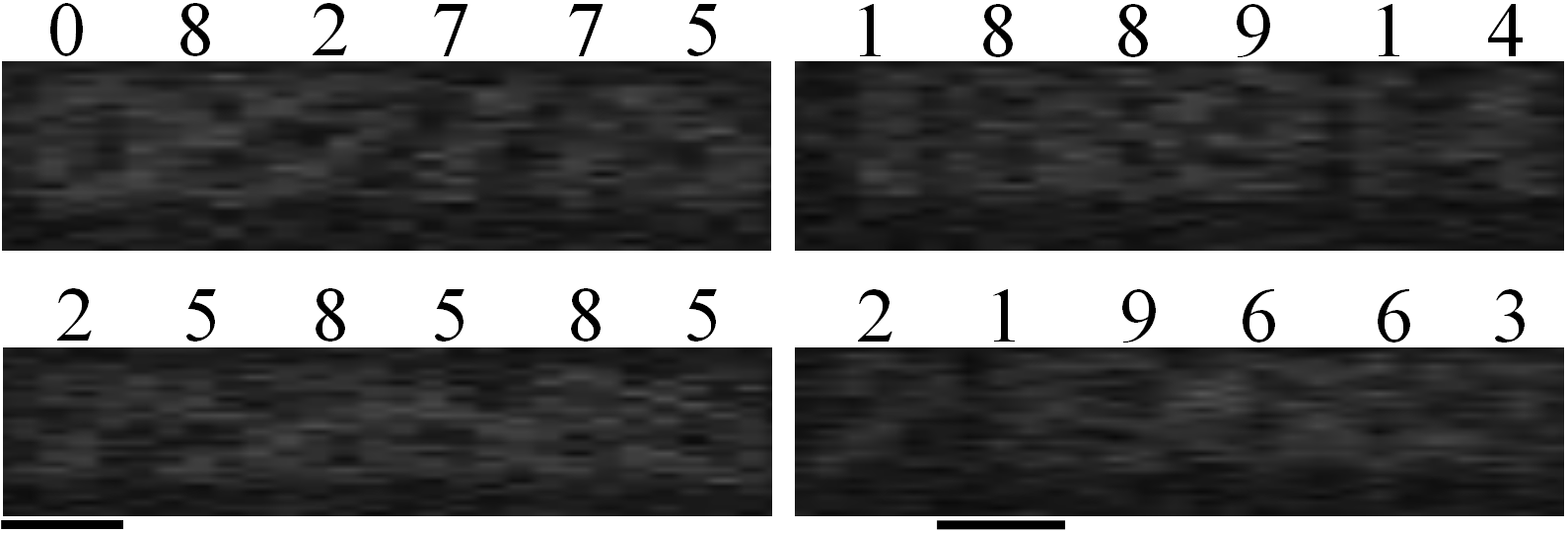}
	\caption{Examples of cropped 6-digit security code. Ground truth labels are shown above each strip, with the underline highlighting the wrong prediction of digits by our classifier. \label{fig:pin_prediction}}
\end{figure}

\begin{table}[t]
\renewcommand{\arraystretch}{1}
\centering
\resizebox{0.8\columnwidth}{!}{
\begin{tabular}{l|ccc}
\toprule
&\textbf{6 digits}&$\mathbf{\geq5}$ \textbf{digits}&$\mathbf{\geq4}$ \textbf{digits}\\
\midrule
\textbf{Acc. (\%)}&50.5&89.5&99.0\\
\bottomrule
\end{tabular}
}
     \caption{Accuracy of predicting partial security code correctly with the CNN classifier in our security code attack.}
    \label{tab:pin_cum}
\end{table}

\section{Experiments} \label{sec:experiments}
In this section, we first conduct experiments on iPhone~6s to analyze the properties of our attack on the basic single-device scenario.
Specifically, we look into the dimensions that can potentially impact the classification performance, such as size and heterogeneity of the training data (Section~\ref{subsec:pin_train}), for further analyzing the attacker's capability in various attack settings.

Then, we test our attack using more phones (iPhone~6-A, iPhone~6-B, and Honor 6X) to validate the effectiveness of our attack in more challenging scenarios, such as cross-device attack, magazine occlusion, and interference from environmental signal noise.
The specifications of different phones can be found in Table~\ref{tab:target_specs}, and the detailed data collection settings are shown in Table~\ref{tab:phonedata}.

\subsection{Security Code Attack}
\label{sec:att}
In our practical security code attack, we use an Apple iPhone~6s as the target device.
We collect 10 sessions of grid data, each of which contains $32000$ emage examples.
Through human inspection, we drop one session due to an obvious data quality issue.
For inter-session evaluation on the grid data, 2 of the remaining 9 valid sessions are fixed as the test
set for all the experiments, where session 8 represents a well-positioned antenna scenario and session 9 is for badly-positioned antenna scenario.
The training set is gradually enlarged by adding more of the remaining sessions.
Specifically, we try four sizes of
training set, which respectively consist of 1, 3, 5 and 7 sessions, denoted as Training 1, Training 2, Training 3 and Training 4, as illustrated in Figure~\ref{fig:sec_batch}.
Each resulting digit emage will be fed as input to train our CNN classifier, following the principles in Section~\ref{subsec:pin_train}.

We simulate 200 text messages, each of which contains a 6-digit security code, making sure they look very close to the real case, as shown in Figure~\ref{fig:screen_display} (right).
In this case, each emage of the security code, with a size of $126 \times 31$, (see Figure~\ref{fig:pin_prediction} for some examples) is evenly divided into six.

The best overall accuracy (89.8\% in Table~\ref{tab:pin_diff_overall}) with respect to all $200\times6=1200$ individual digits is achieved when using all of the 7 training sessions (more details about the impact of training data amount will be discussed in Section~\ref{subsec:pin_train}).
As can be seen in Table~\ref{tab:pin_diff_overall}, the accuracy differs for different digits, with the highest (99.1\%) achieved for digit 4, and lowest (75.8\%) for digit 3.
Figure~\ref{fig:pin_prediction} shows some examples for the security code along with the ground truth and prediction results.
It demonstrates that our approach can correctly predict the digits with high accuracy although the digits are hardly recognizable to the human eye.

\begin{figure}[t]
	\centering

	\includegraphics[width=0.8\linewidth]{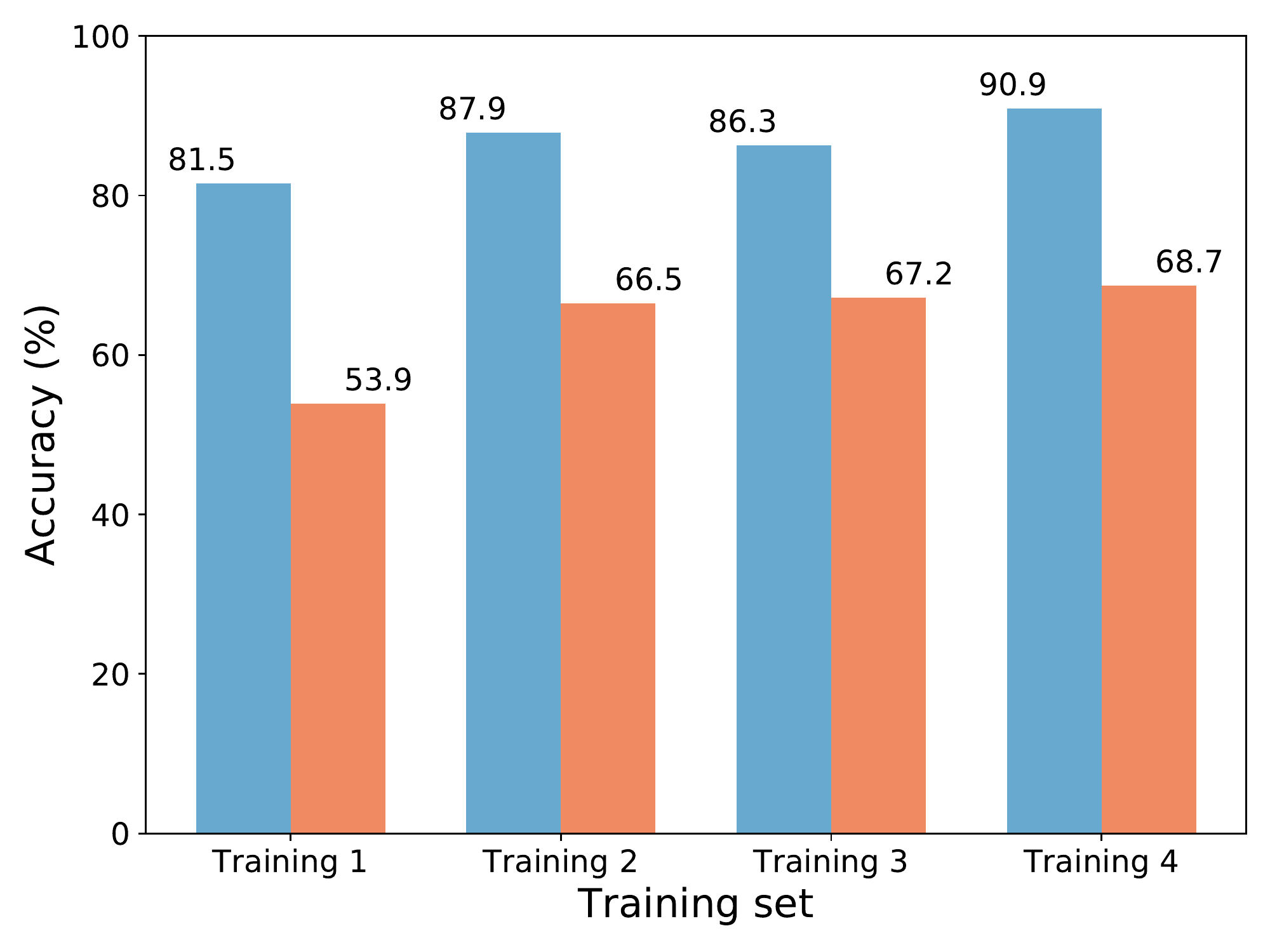}
	\caption{Inter-session accuracy (grid data) of our security code attack for different training sets with gradually increased size. The two bars for each training set represent two different test sessions (session 8 and 9).\label{fig:pin_cross_sess_acc}}
	\vspace{-0.5cm}
\end{figure}

In practice, attackers may have various query budgets for fully uncovering the security code (with all the 6 digits being correct).
So, in Table~\ref{tab:pin_cum}, we present the accuracy results when four security code digits or more can be correctly predicted by our classifier.
It can be observed that, with one attempt, the attacker can fully recognize the security code at 50\% of the cases.
The probability of recognizing four or more digits can reach 99\%, showing that our approach can present a serious threat in practice.

\subsection{Data Analysis on Grid Data}
\label{subsec:pin_train}

We first consider the scenario where the attacker can train the classifier on data sampled from the same distribution as the attacked security code.
This can be regarded as the best-case scenario, although almost impossible in most practical cases.
Specifically, we achieve an accuracy of 86.5\% within the session used in Training 1.

\begin{figure}[!t]
	\centering

	\includegraphics[width=0.8\linewidth]{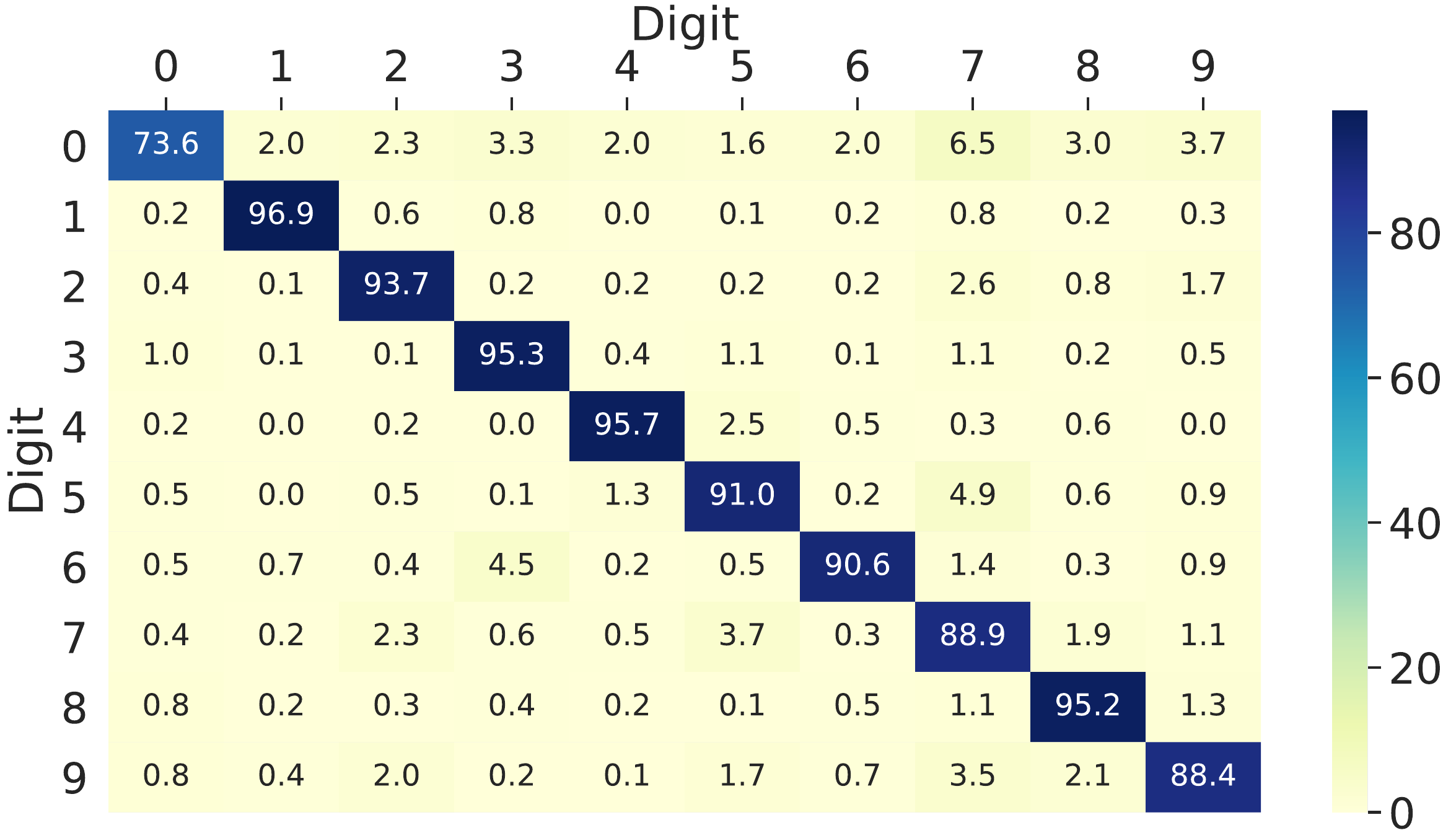}
	\caption{Confusion matrix of the inter-session accuracy (grid data) in our security code task. Results are from the classifier trained on Training 4 and tested on session 8.\label{fig:pin_conf_class}}
\end{figure}

Inter-session evaluation represents a more realistic attack scenario, where the training data from the same session of the target is not accessible, but the attacker can simulate similar data using the same settings.

Figure~\ref{fig:pin_cross_sess_acc} shows the inter-session accuracy of the four classifiers trained on different training sets: training 1, 2, 3 and 4.
It can be observed that the accuracy improves as we increase the number of training sessions.
We can also observe that inter-session accuracy with only one training session is lower than the multi-crop grid case.
However, using more training data with multiple sessions could alleviate this issue, leading to a high accuracy of 90.9\% for Training 4 (with seven training sessions).
This validates our assumption that incorporating heterogeneous sessions could help alleviate the impact of the random noise introduced to the emage generation.
One detailed classification result with respect to different classes are shown in the confusion matrix in Figure~\ref{fig:pin_conf_class}.
We also notice that there is a difference between the prediction performance between two test sessions, which might be explained by their different data quality.

\begin{figure}[!t]
\begin{minipage}[b]{0.48\columnwidth}
\centering
\includegraphics[width=1\columnwidth,height=1\columnwidth]{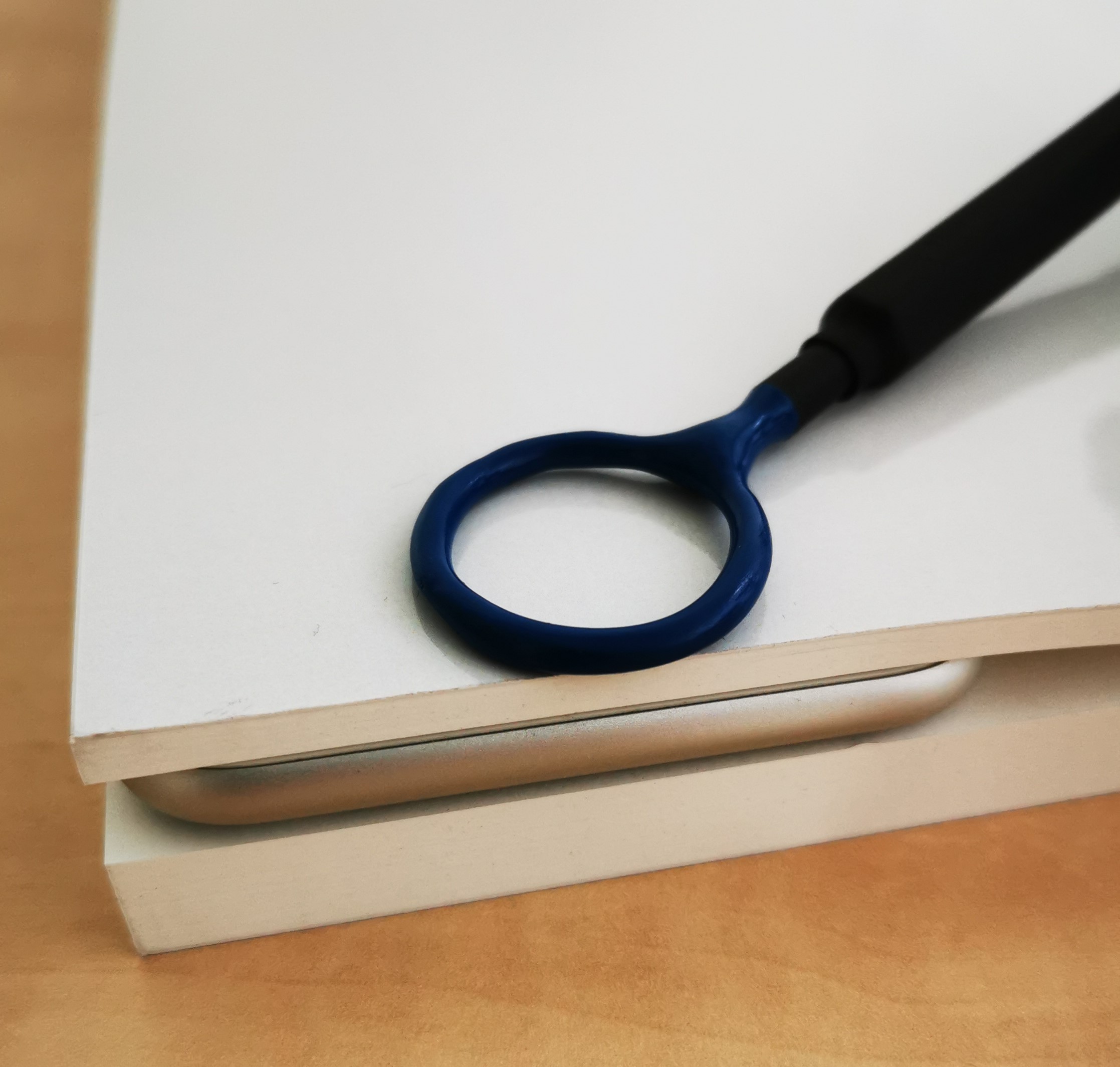}
\end{minipage}
\hspace{0.2cm}
\begin{minipage}[b]{0.48\columnwidth}
\centering
\includegraphics[width=1\columnwidth,height=1\columnwidth]{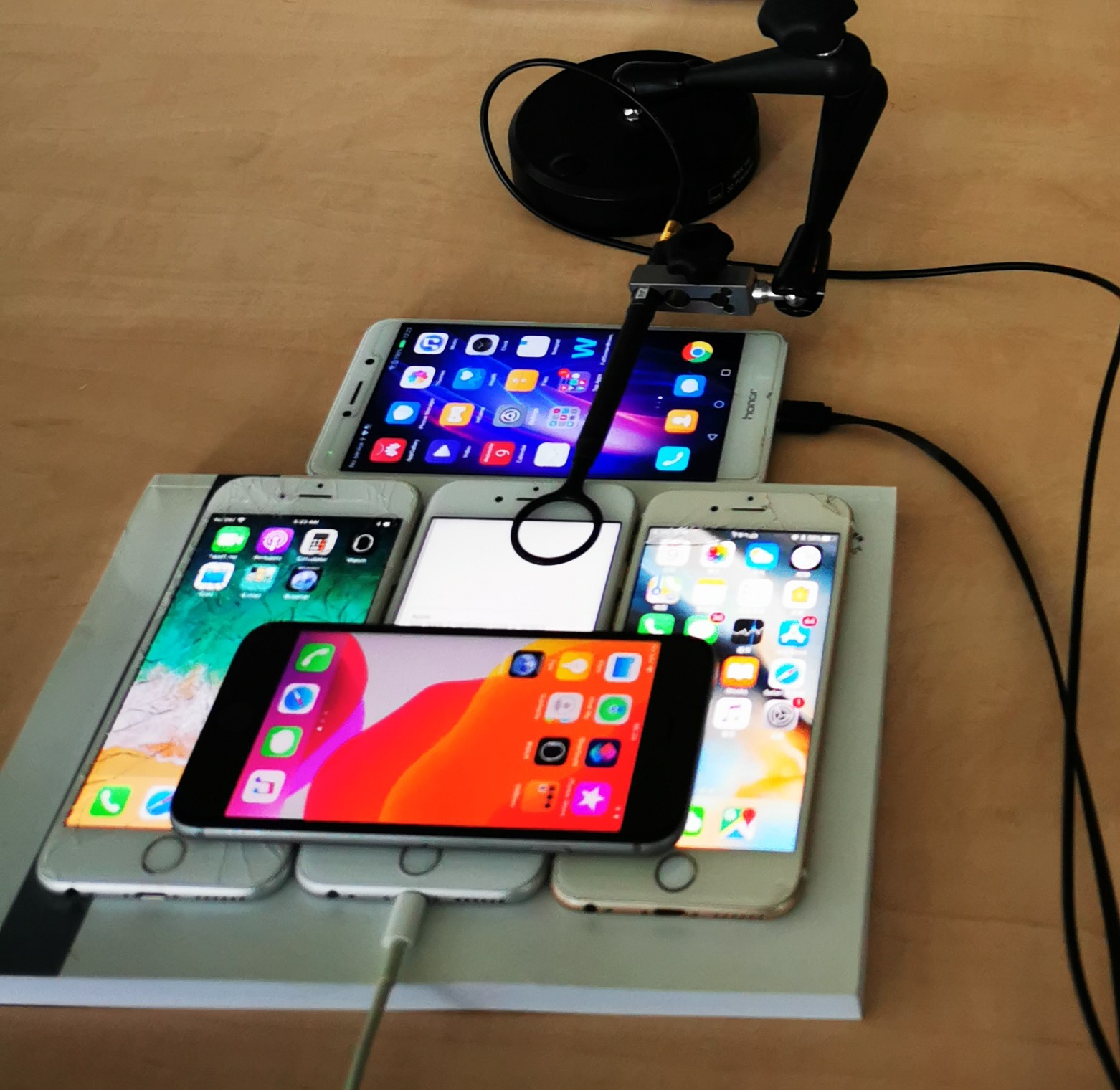}
\end{minipage}
\caption{Pictures of the Magazine setting (left), with the phone in between the magazine pages and the probe on top, and the With Noise setting (right).}
\label{fig:exp_setup}
\end{figure}

\begin{table*}[!t]
\centering
	\begin{tabular}{p{2cm}|ccccccc}
		\toprule

		\textbf{Acc. (\%)}& \makecell{\textbf{Test} \\ \textbf{(Grid)}}   &\makecell{\textbf{Single-device-1} \\ \textbf{(Security code)}} &\makecell{\textbf{Single-device-2} \\ \textbf{(Security code)}} &\multicolumn{2}{c}{\makecell{\textbf{Magazine}\\ \textbf{(Security code)}}} &\makecell{\textbf{With Noise} \\ \textbf{(Security code)}}&\makecell{\textbf{iPhone~6-B} \\ \textbf{(Security code)}}\\
		&&&&70 pages&200 pages\\
		\midrule
		iPhone~6-A   & 73.42 & 41.42 & 47.08 &14.38 &-&63.29&61.54 \\
		Honor~6X & 94.38 & 74.00 &74.00&-&65.79&64.25&-\\

		\bottomrule
	\end{tabular}

    \caption{Inter-session classification of our security code attack for different phones and different test settings. Grid means the multi-crop grid test data is used, and Security Code means the simulated text message test data is used.
    The training set stays the same in all test settings for each device.
    Single-device-1 and Single-device-2 refer to two different test sessions.}
	\label{tab:phacc}
\end{table*}

\subsection{Experiments on Other Phones}
In this section, we conduct experiments on different phones to further validate the general effectiveness of our security code recognition on different devices. We show the potential of the recognition in more challenging and realistic scenarios, including cross-device attack, antenna occlusion by a magazine, and interference from the signal noise generated by surrounding phones (cf. Figure~\ref{fig:exp_setup}).
The cross-device attack consists of training the recognition algorithm on the data from one device and testing the model on data from another unit of the same model.
Specifically, we use two iPhone~6, namely, iPhone~6-A and iPhone~6-B, and make sure that they have the same version of the iOS system, and not refurbished.
Five sessions of data are collected for training the recognition model on iPhone~6-A, and two test sessions of security code data are collected for testing.
Additionally, we collect a session of testing data with the antenna occluded by a magazine,  another test session from iPhone~6-B and a test session with background noise.
The measurement setups for occluding the antenna and simulating the background noise are shown respectively in Figure~\ref{fig:exp_setup}.
Each of the above four testing sessions contains 200 different security codes and for each code, we repeat the frame twice for a more stable recognition.
Our attack can also work on a refurbished iPhone (iPhone 6-C, see Table~\ref{tab:target_specs}), but no quantitative results are reported in order to maintain fair comparison.

Table~\ref{tab:phacc} summarizes our experimental results under different test settings corresponding to the data descriptions in Table~\ref{tab:phonedata}, i.e., Multi-crop, Single-device, Magazine, Noise, and Cross-device.
As can be seen, our model achieves high accuracy for the original multi-crop data.
For other settings, as expected, the performance drops due to the generalization gap but still being effective enough in most cases.
Specifically, the high cross-device accuracy suggests the effectiveness of our attack in a more realistic scenario, where the device used for collecting the training data is not necessarily the target device.
The results on an Android phone, Honor 6X, with four sessions of training data, verify that the effectiveness of our attack is not limited to the specific phone type, iPhone.
we can also observe that the single-device and cross-device sessions of the security code yield different prediction performance, which might be explained by their different data quality, as also reported for iPhone 6s (cf. Section~\ref{sec:att}).

For the magazine setting, the accuracy drop can be explained by the signal strength of the antenna. The magnetic probe can be considered as a magnetic dipole, for which it holds that the power density is dependent on a factor $r^{-5}$, for which $r$ is the distance between the probe and the origin of radiation~\cite{Ida2000}. Therefore, placing the magnetic probe a little bit further away from the origin of radiation, already has some significant consequences on the quality of the received signal.
Specifically, we find the performance of iPhone 6-A drops dramatically with 70 pages, but for Honor 6X, the performance is better maintained even with a thicker magazine of 200 pages because of its higher leakage of signals.
The high single-device accuracy (74\% for both) also confirms this higher signal leakage of this Honor 6X phone than iPhone~6.

We also find that our attack can work on the OLED screen by conducting a preliminary exploration of Samsung Galaxy A3 (2015).
However, since this phone is disassembled, we do not go further for quantitative details.

\subsection{Discussion}

\begin{figure}[t]
	\centering

	\includegraphics[width=0.8\columnwidth]{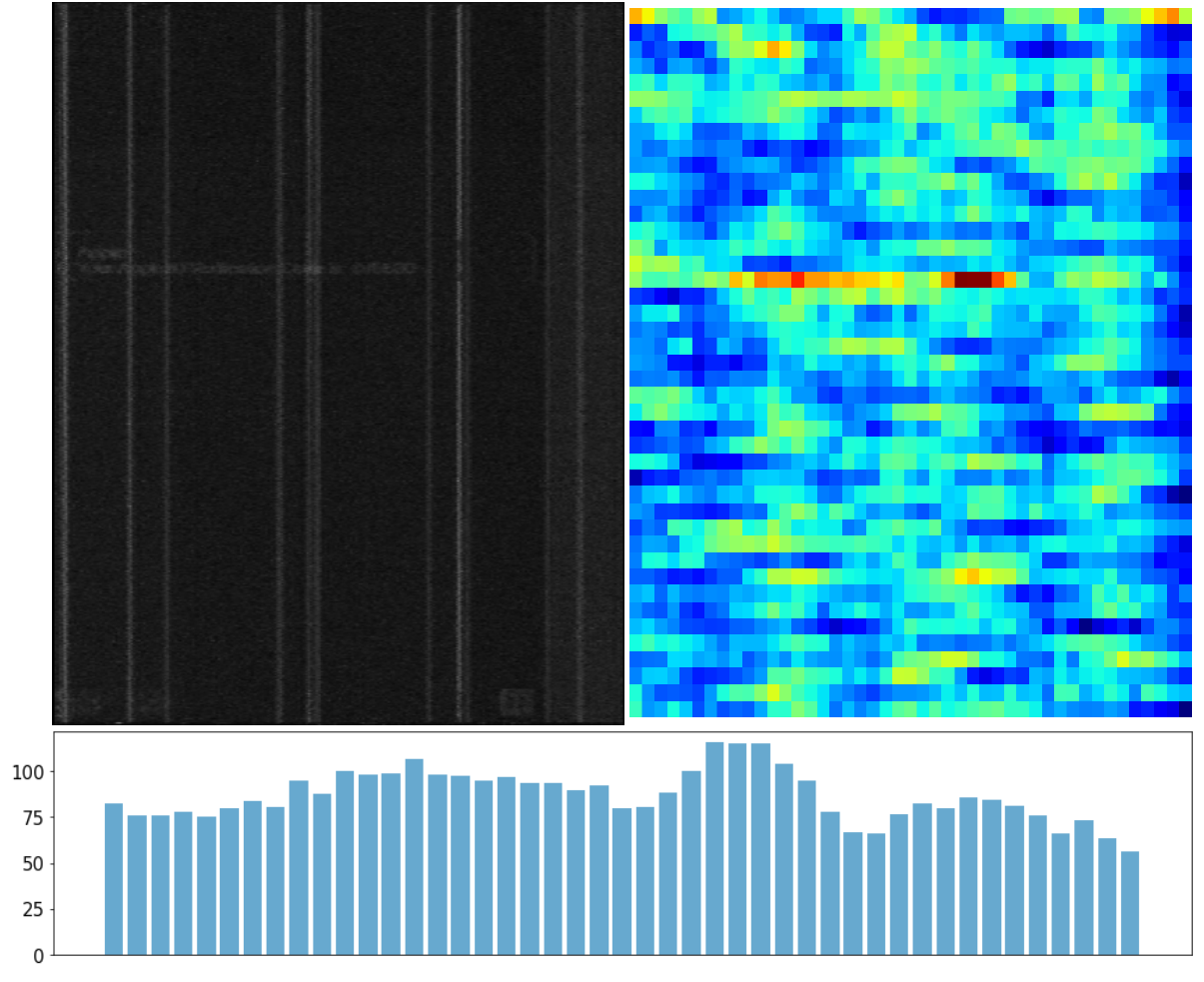}
	\caption{Top: An emage and its predicted activation map by the pre-trained model on iPhone~6s. Warmer color represents higher prediction confidence. Bottom: Activation responses in the row of the text message}
	\label{fig:slidingwindow}
\end{figure}

\begin{figure*}[!t]
\vspace{-1cm}
  \centering
  \begin{subfigure}{.09\textwidth}
    \centering
    \includegraphics[width=\linewidth]{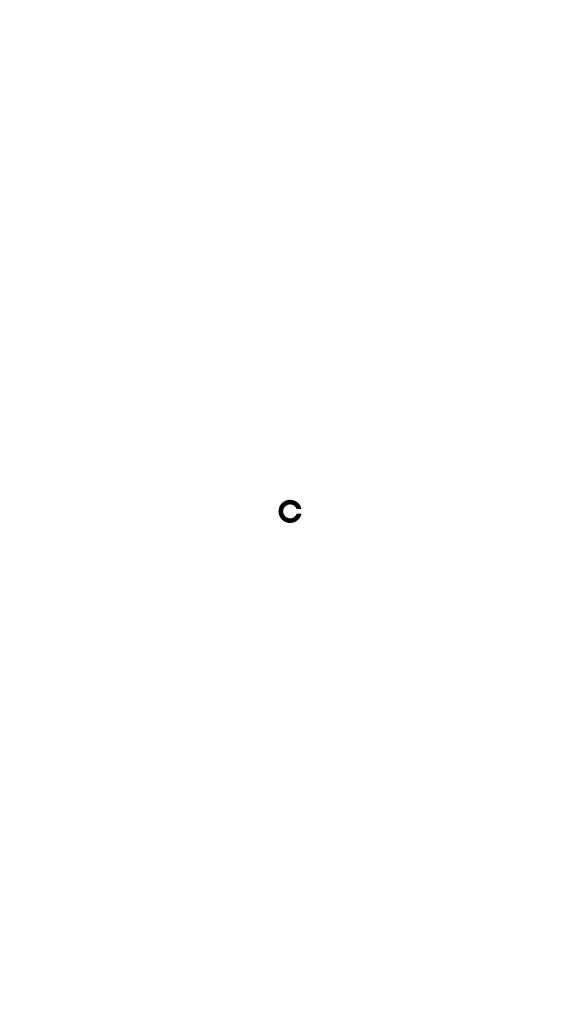}
    \label{fig:letterc0_appendix}
  \end{subfigure}%
  \begin{subfigure}{.09\textwidth}
    \centering
    \includegraphics[width=\linewidth]{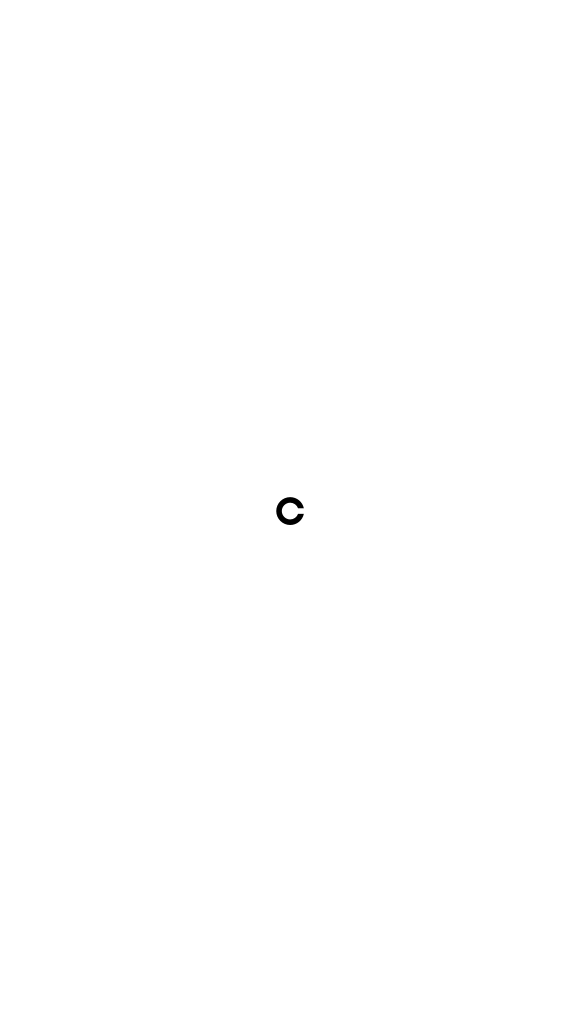}
    \label{fig:letterc1_appendix}
  \end{subfigure}%
  \begin{subfigure}{.09\textwidth}
    \centering
    \includegraphics[width=\linewidth]{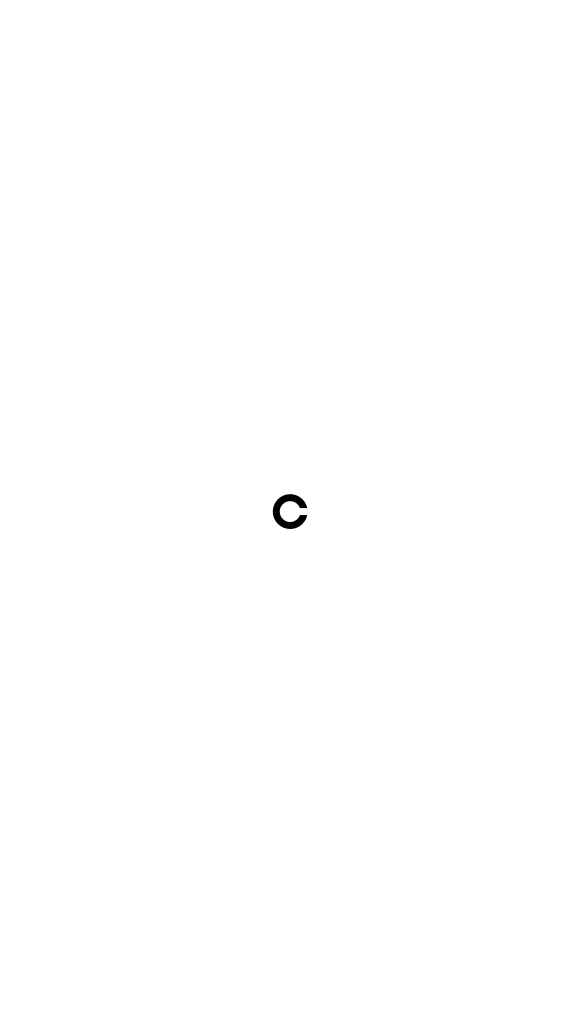}
    \label{fig:letterc2_appendix}
  \end{subfigure}%
  \begin{subfigure}{.09\textwidth}
    \centering
    \includegraphics[width=\linewidth]{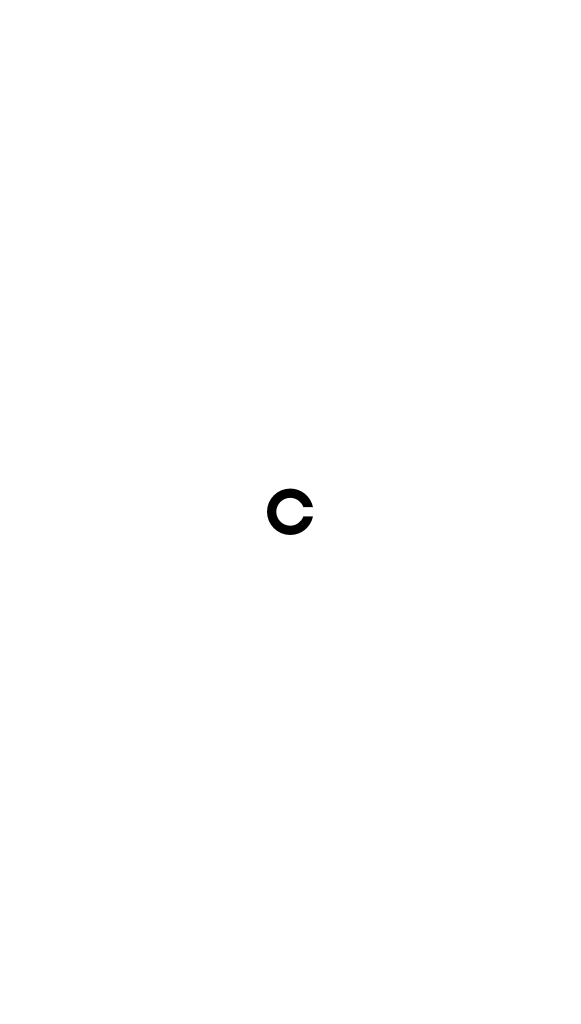}
    \label{fig:letterc3_appendix}
  \end{subfigure}%
  \begin{subfigure}{.09\textwidth}
    \centering
    \includegraphics[width=\linewidth]{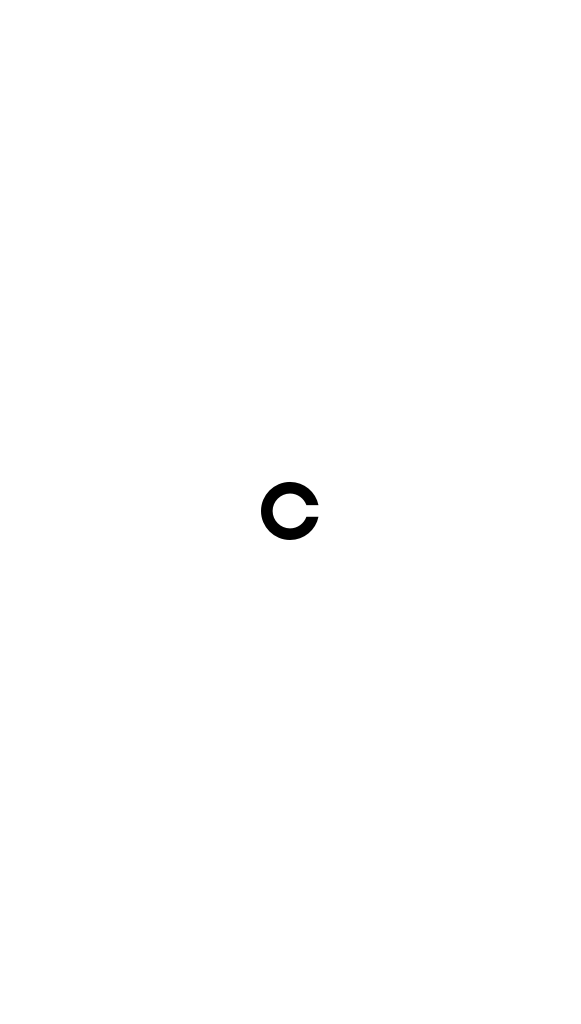}
    \label{fig:letterc4_appendix}
  \end{subfigure}%
  \begin{subfigure}{.09\textwidth}
    \centering
    \includegraphics[width=\linewidth]{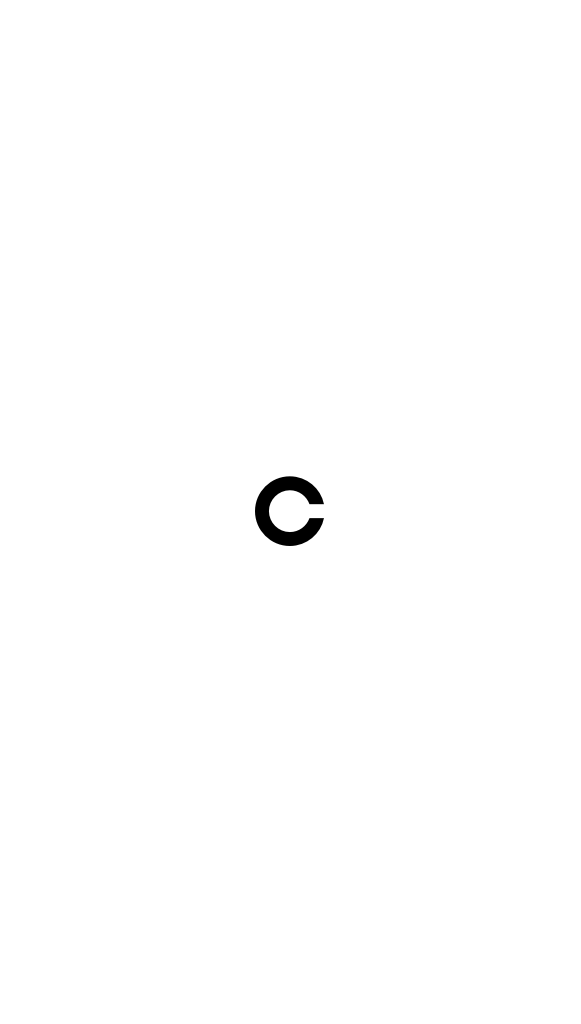}
    \label{fig:letterc5_appendix}
  \end{subfigure}%
  \begin{subfigure}{.09\textwidth}
    \centering
    \includegraphics[width=\linewidth]{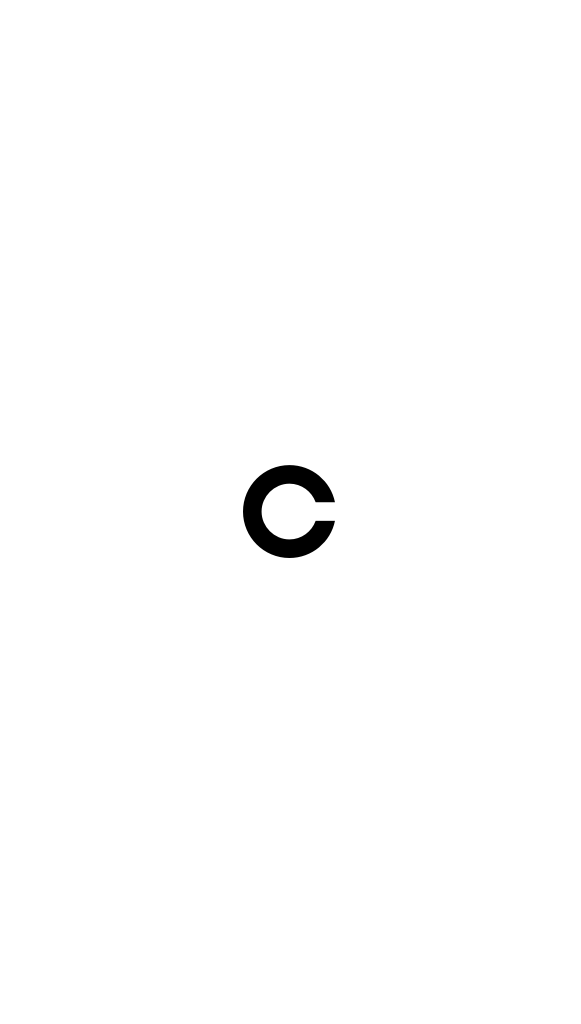}
    \label{fig:letterc6_appendix}
  \end{subfigure}%
  \begin{subfigure}{.09\textwidth}
    \centering
    \includegraphics[width=\linewidth]{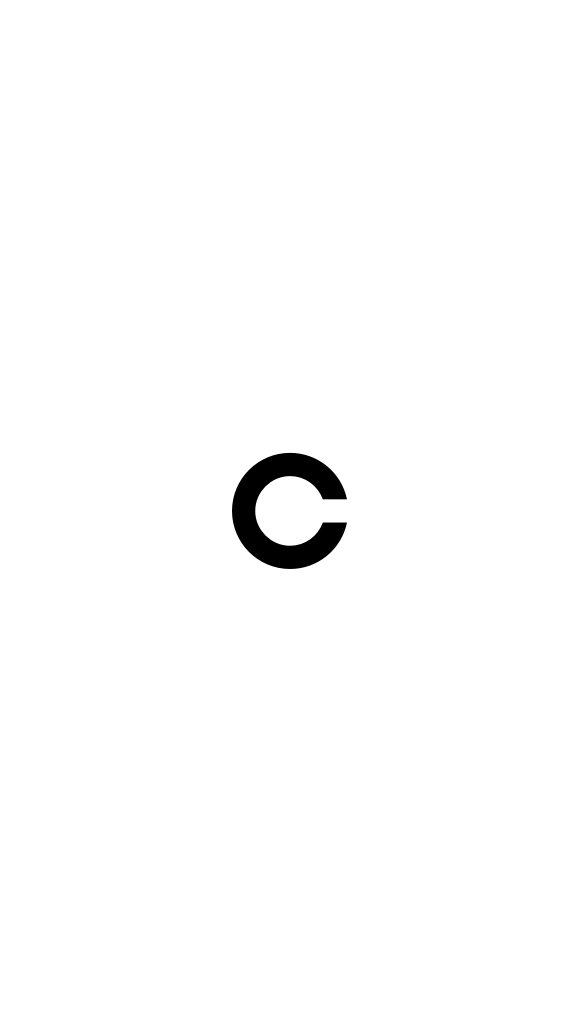}
    \label{fig:letterc7_appendix}
  \end{subfigure}%
  \begin{subfigure}{.09\textwidth}
    \centering
    \includegraphics[width=\linewidth]{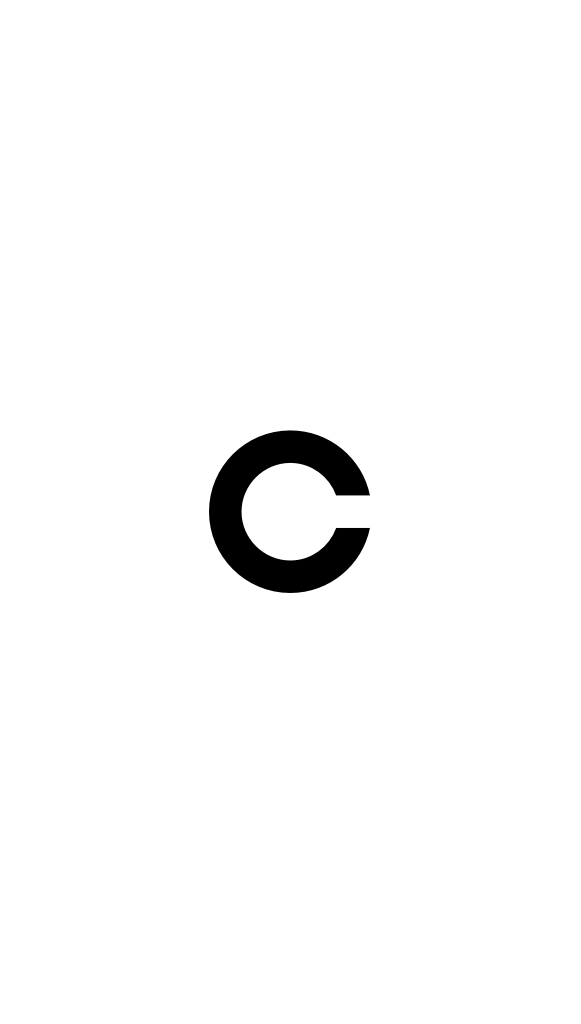}
    \label{fig:letterc8_appendix}
  \end{subfigure}%
  \begin{subfigure}{.09\textwidth}
    \centering
    \includegraphics[width=\linewidth]{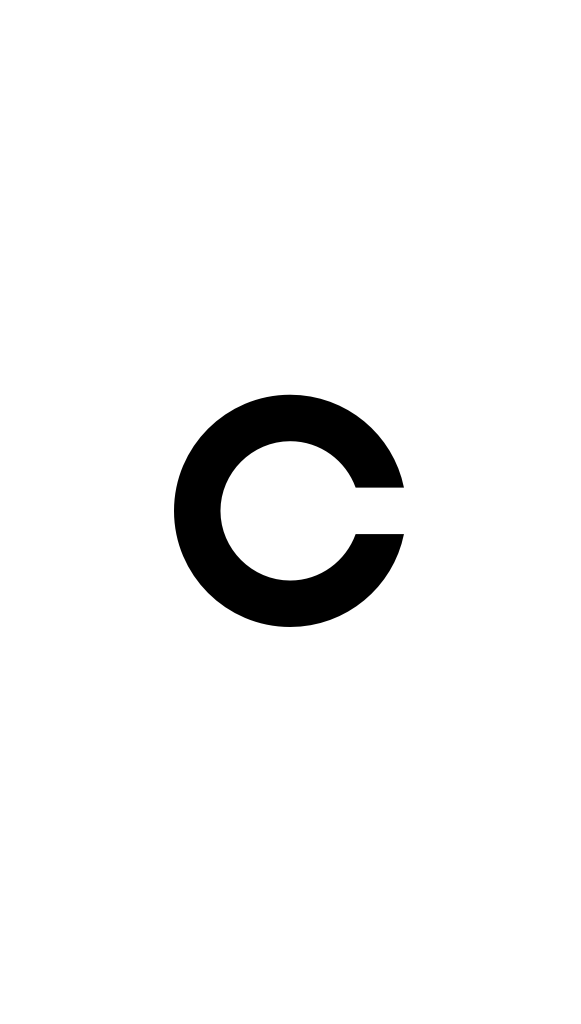}
    \label{fig:letterc9_appendix}
  \end{subfigure}%
  \begin{subfigure}{.09\textwidth}
    \centering
    \includegraphics[width=\linewidth]{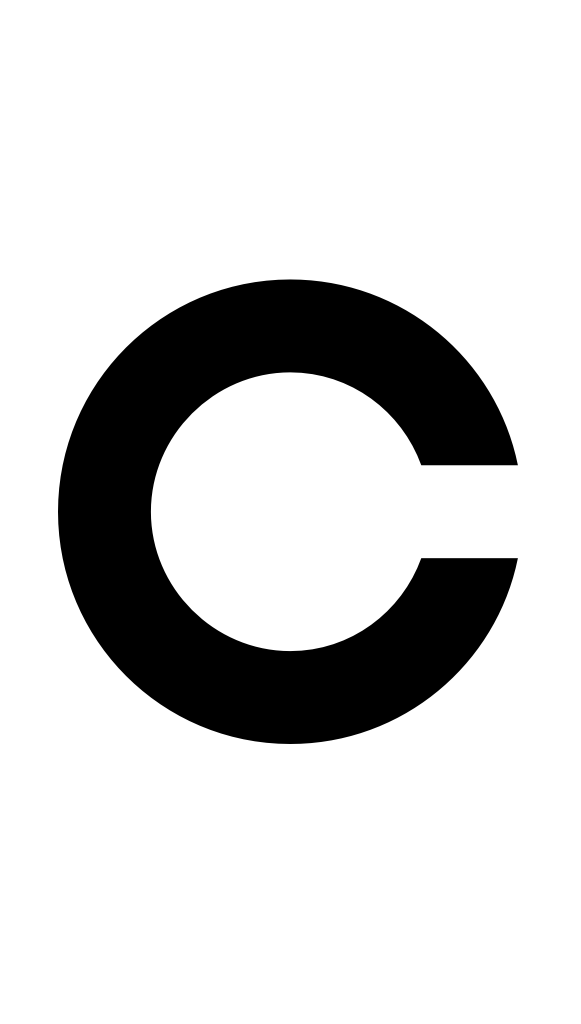}
    \label{fig:letterc10_appendix}
  \end{subfigure}%

  \begin{subfigure}{.09\textwidth}
    \centering
    \includegraphics[width=.98\linewidth]{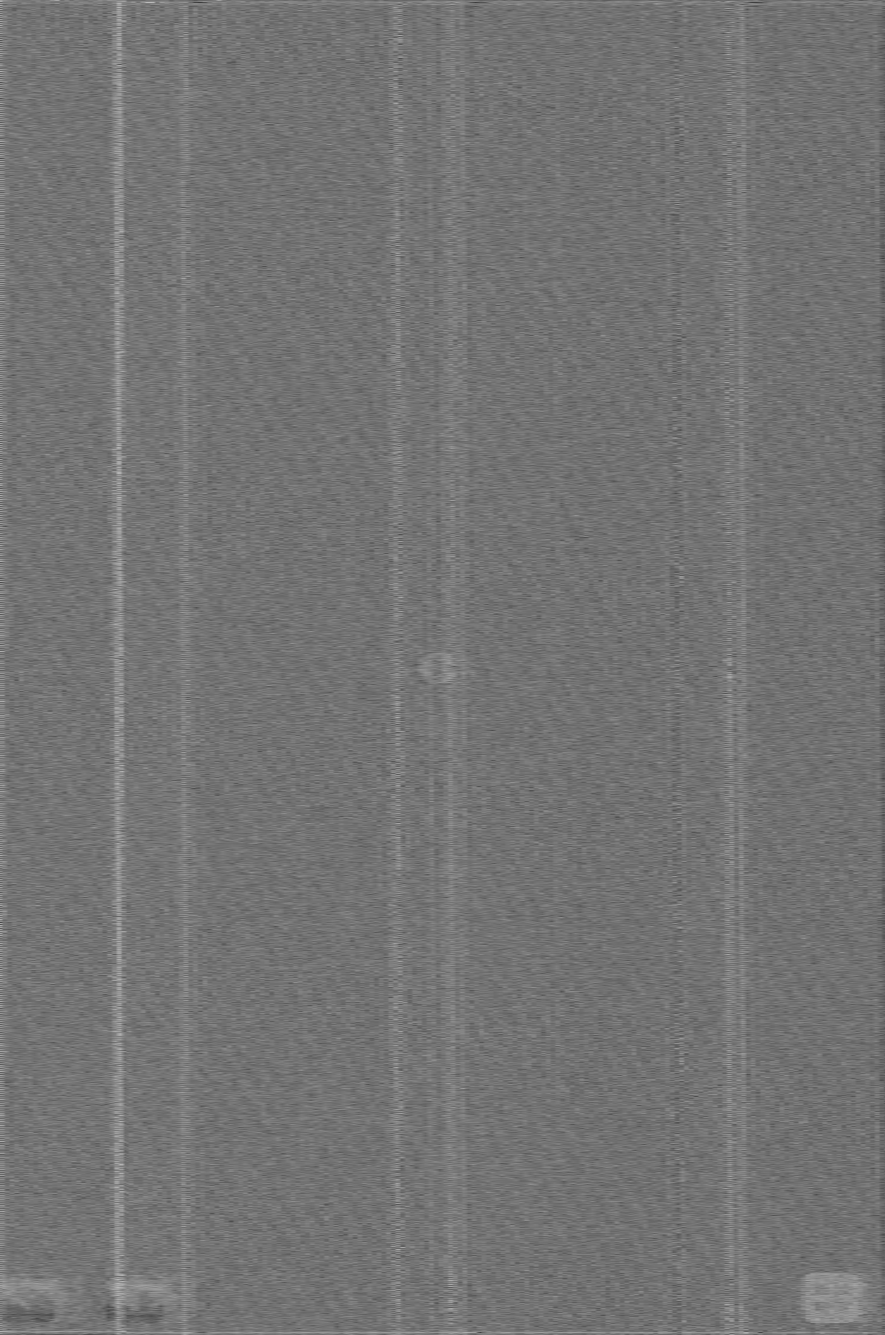}
    \caption{1x}
    \label{fig:emagec0_appendix}
  \end{subfigure}%
  \begin{subfigure}{.09\textwidth}
    \centering
    \includegraphics[width=.98\linewidth]{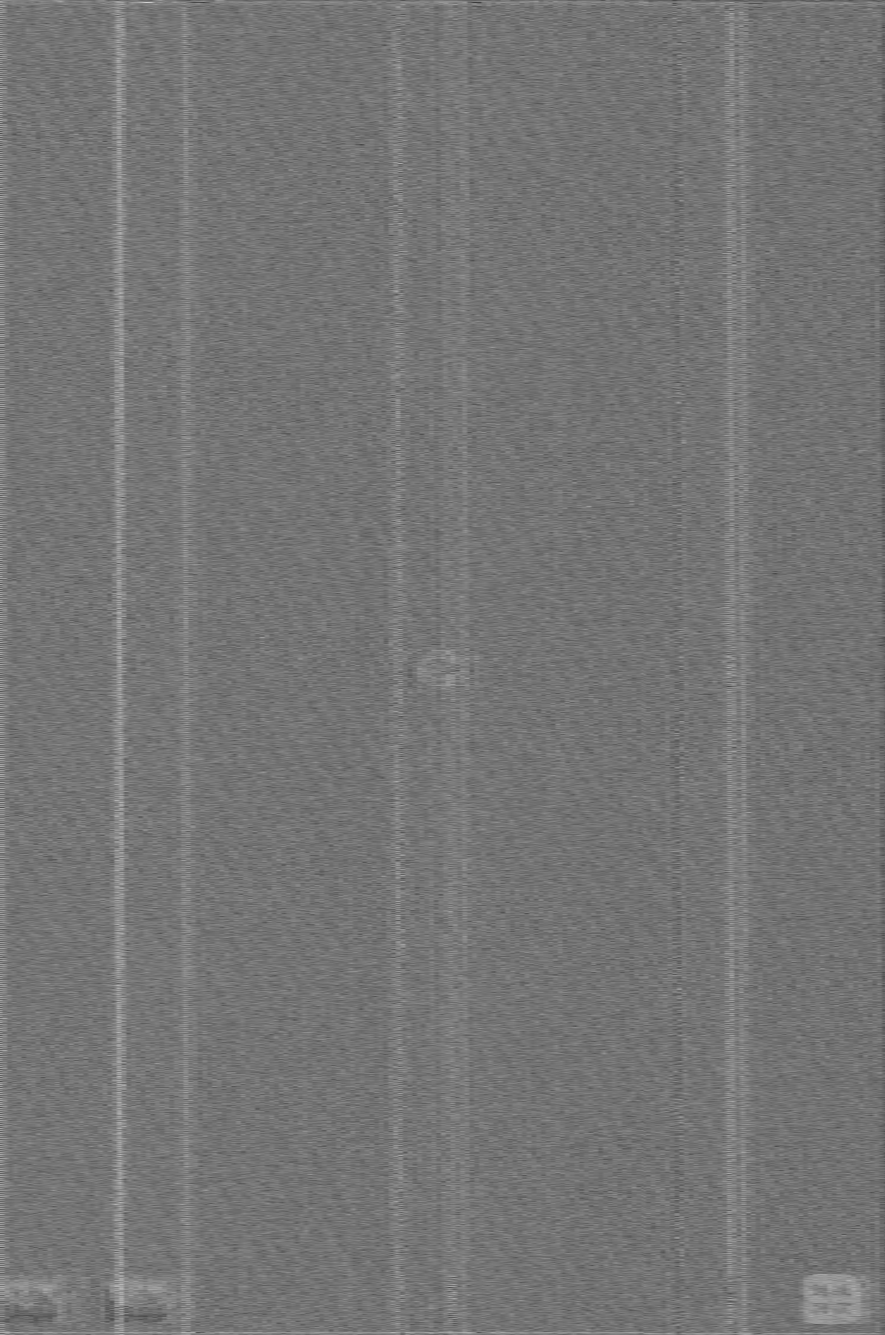}
    \caption{1.2x}
    \label{fig:emagec1_appendix}
  \end{subfigure}%
  \begin{subfigure}{.09\textwidth}
    \centering
    \includegraphics[width=.98\linewidth]{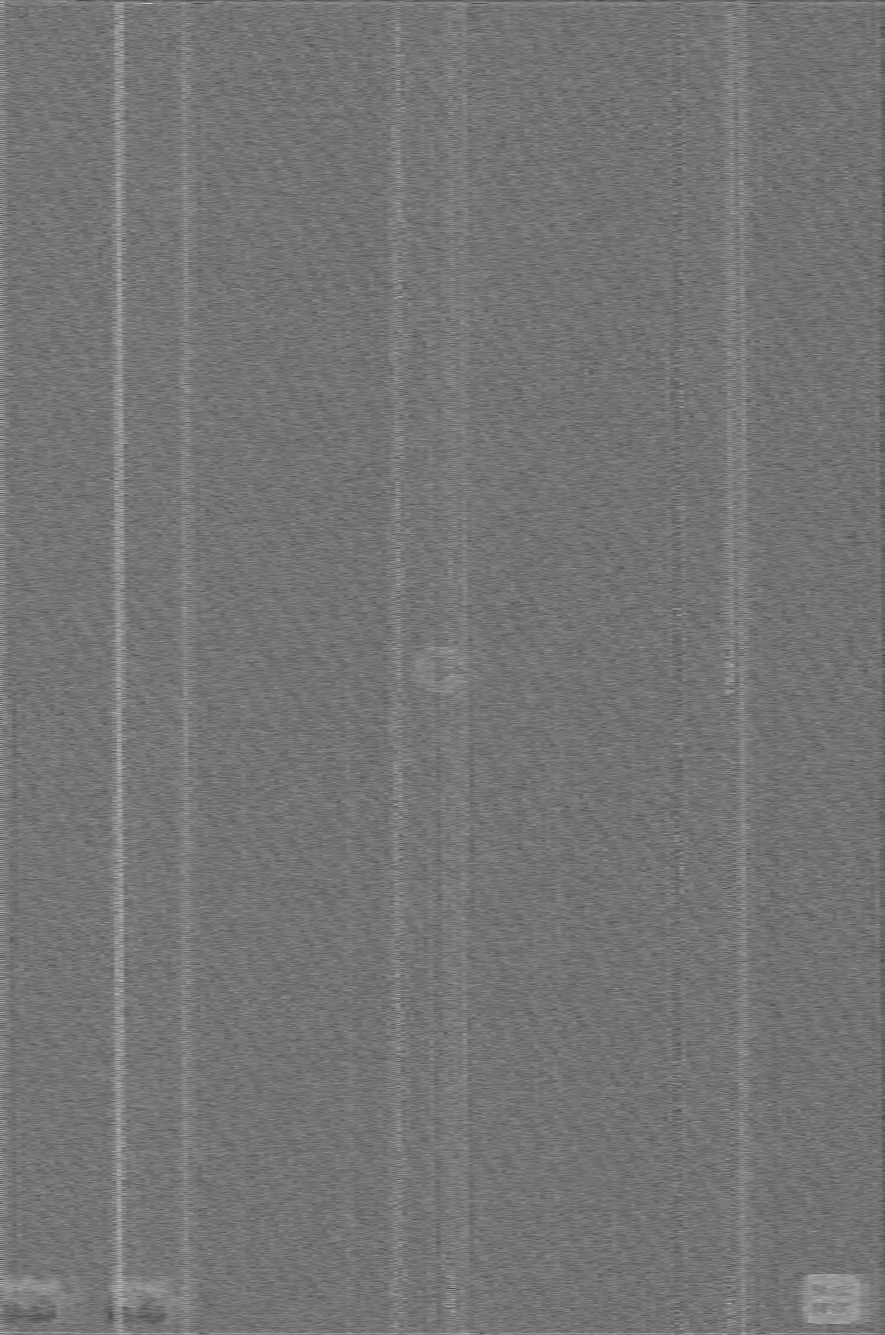}
    \caption{1.5x}
    \label{fig:emagec2_appendix}
  \end{subfigure}%
  \begin{subfigure}{.09\textwidth}
    \centering
    \includegraphics[width=.98\linewidth]{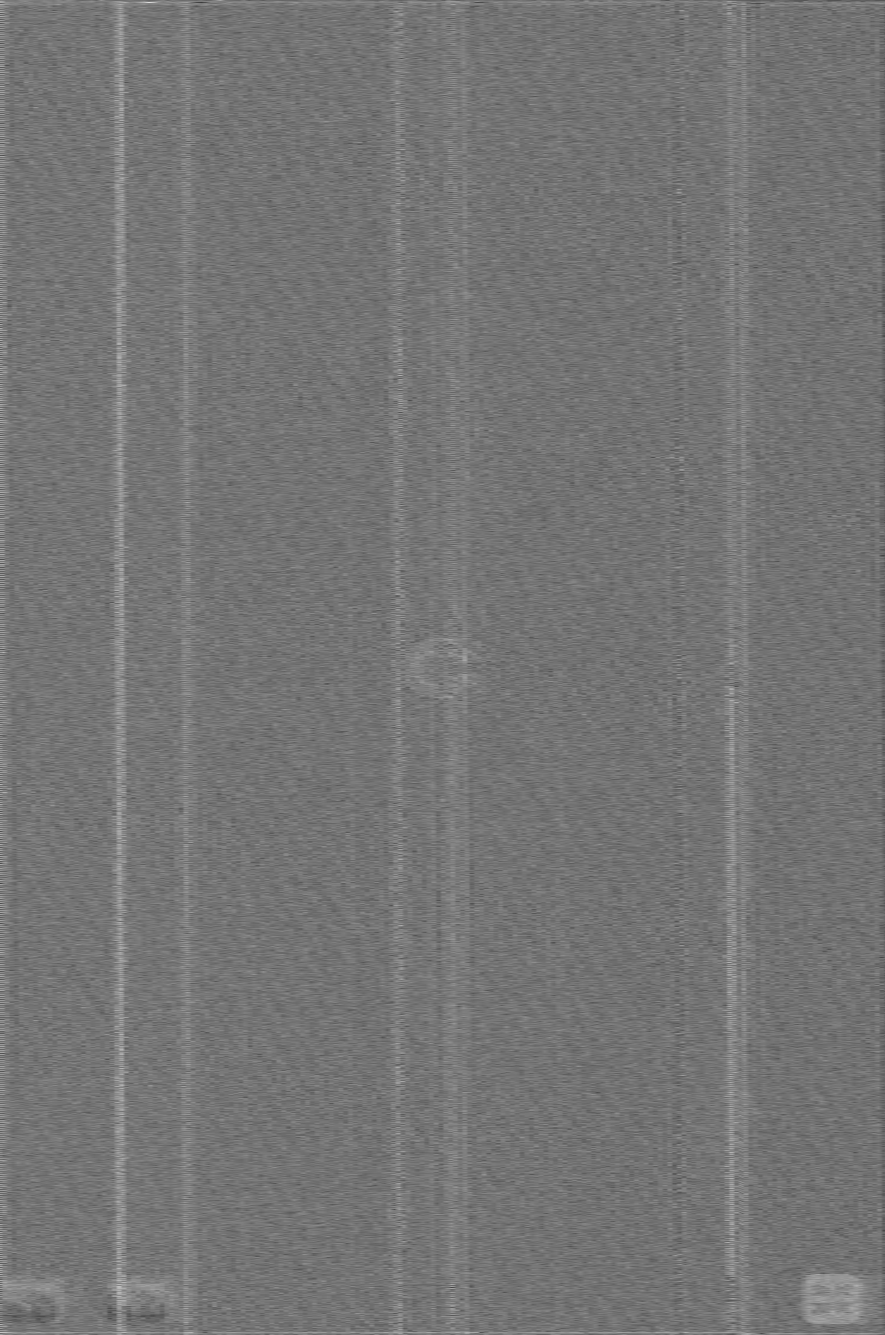}
    \caption{2x}
    \label{fig:emagec3_appendix}
  \end{subfigure}%
  \begin{subfigure}{.09\textwidth}
    \centering
    \includegraphics[width=.98\linewidth]{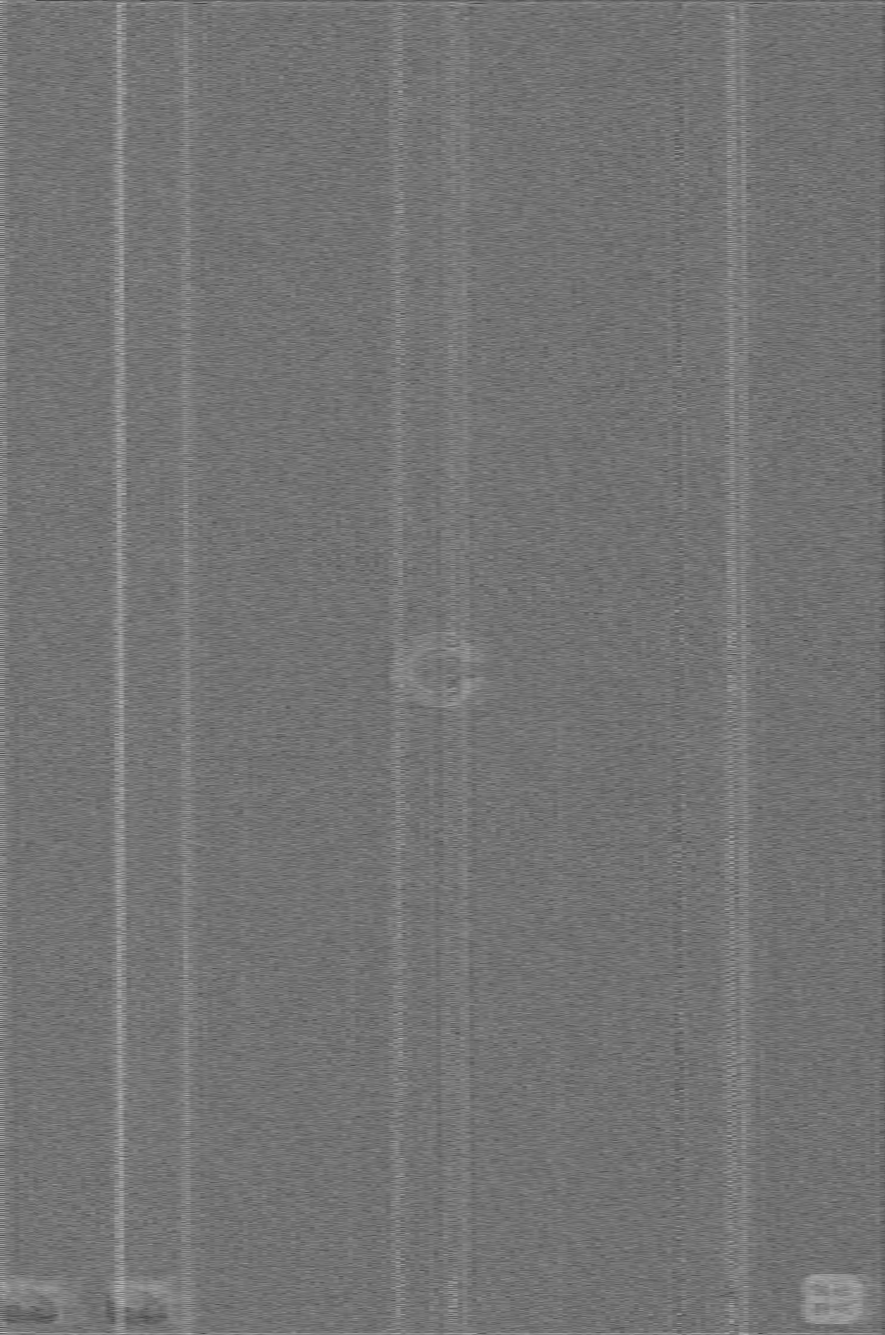}
    \caption{2.5x}
    \label{fig:emagec4_appendix}
  \end{subfigure}%
  \begin{subfigure}{.09\textwidth}
    \centering
    \includegraphics[width=.98\linewidth]{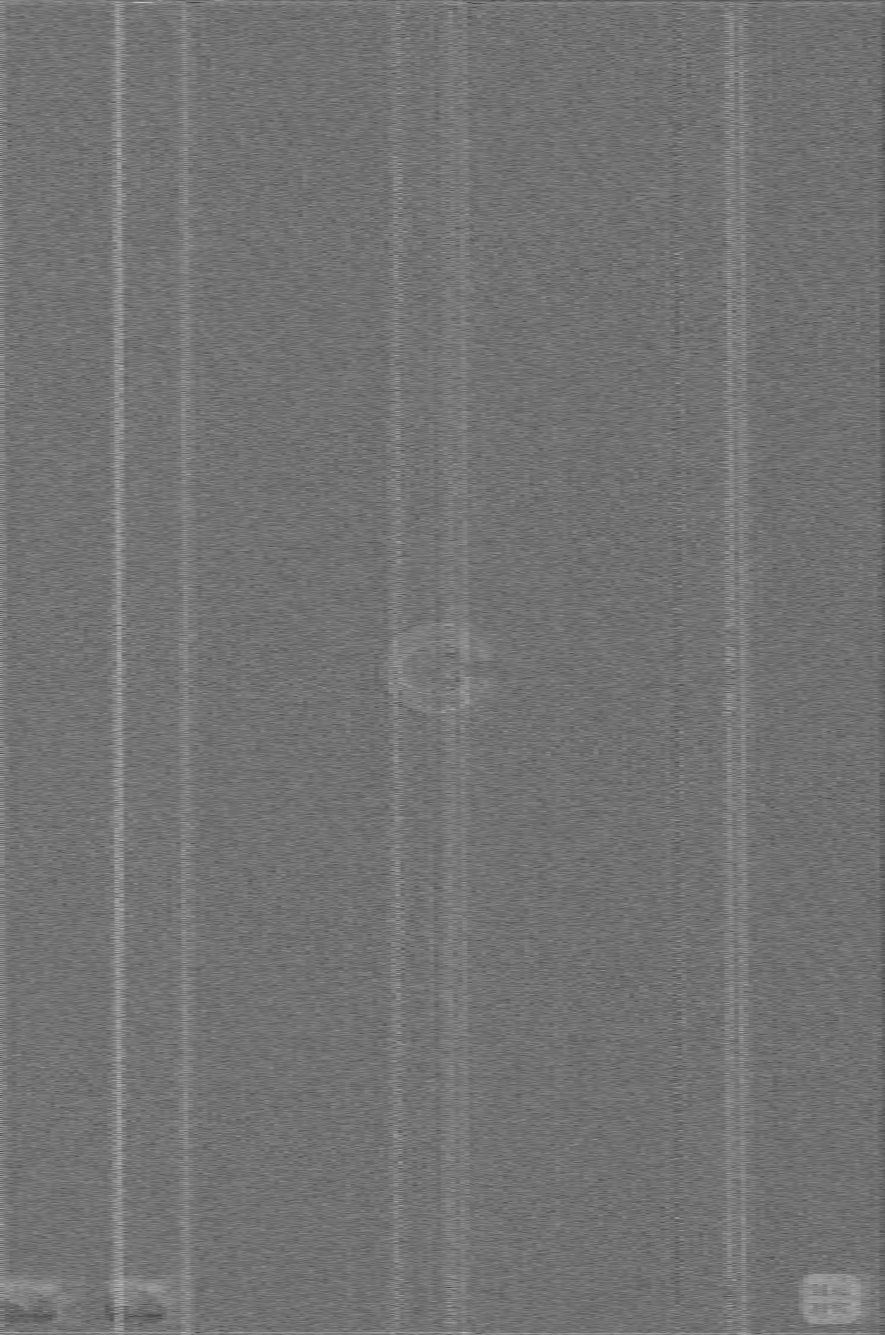}
    \caption{3x}
    \label{fig:emagec5_appendix}
  \end{subfigure}%
  \begin{subfigure}{.09\textwidth}
    \centering
    \includegraphics[width=.98\linewidth]{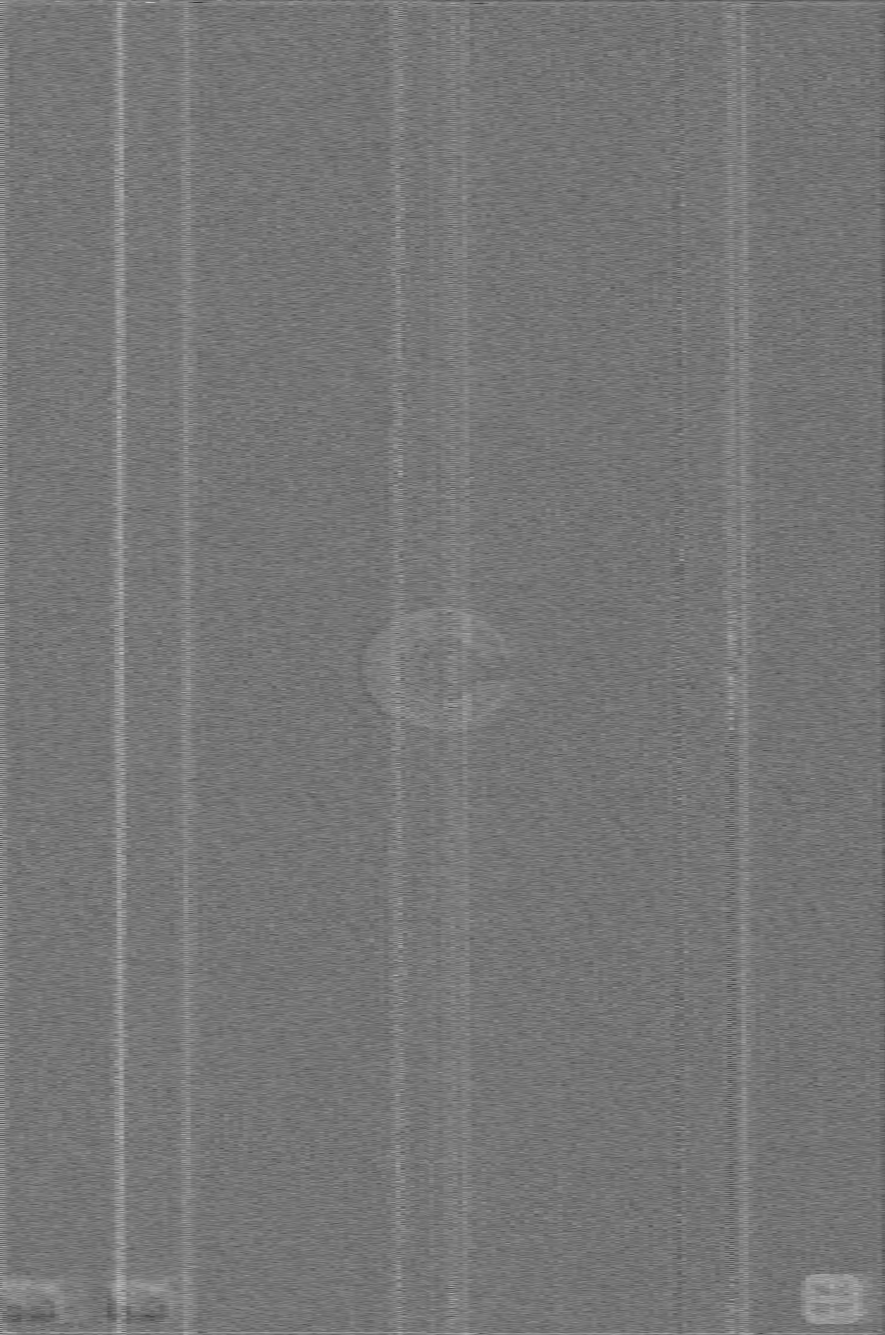}
    \caption{4x}
    \label{fig:emagec6_appendix}
  \end{subfigure}%
  \begin{subfigure}{.09\textwidth}
    \centering
    \includegraphics[width=.98\linewidth]{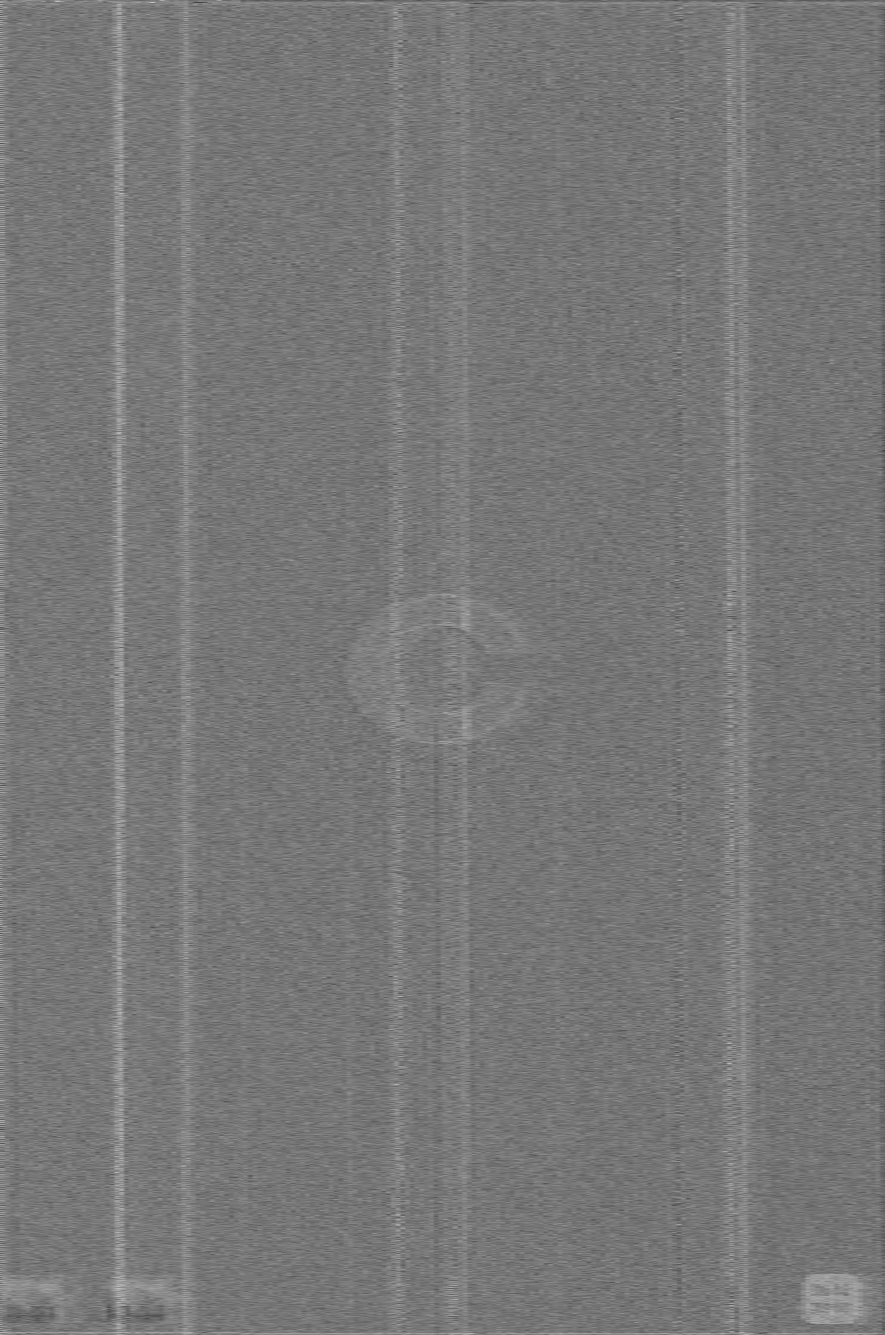}
    \caption{5x}
    \label{fig:emagec7_appendix}
  \end{subfigure}%
  \begin{subfigure}{.09\textwidth}
    \centering
    \includegraphics[width=.98\linewidth]{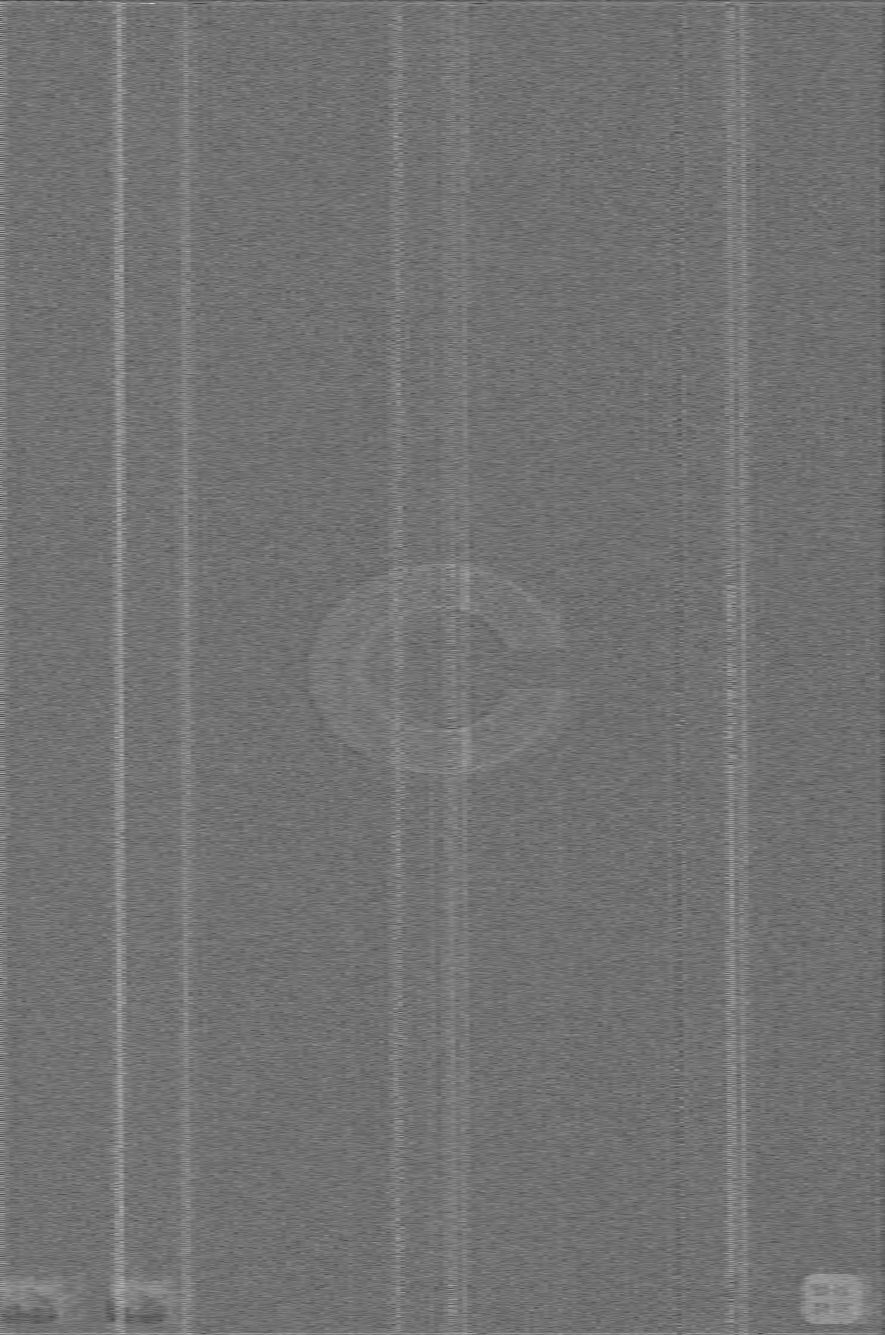}
    \caption{7x}
    \label{fig:emagec8_appendix}
  \end{subfigure}%
  \begin{subfigure}{.09\textwidth}
    \centering
    \includegraphics[width=.98\linewidth]{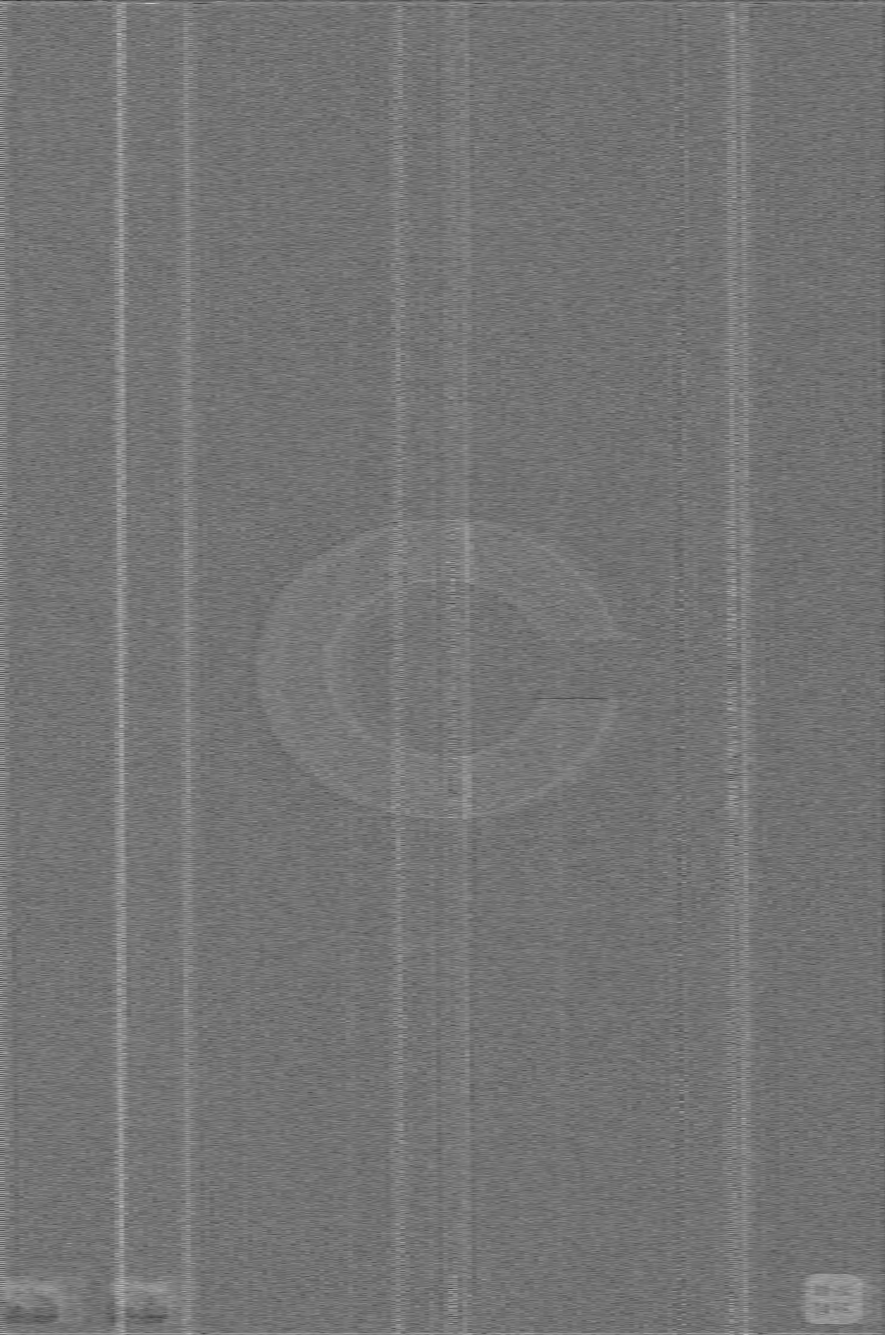}
    \caption{10x}
    \label{fig:emagec9_appendix}
  \end{subfigure}%
  \begin{subfigure}{.09\textwidth}
    \centering
    \includegraphics[width=.98\linewidth]{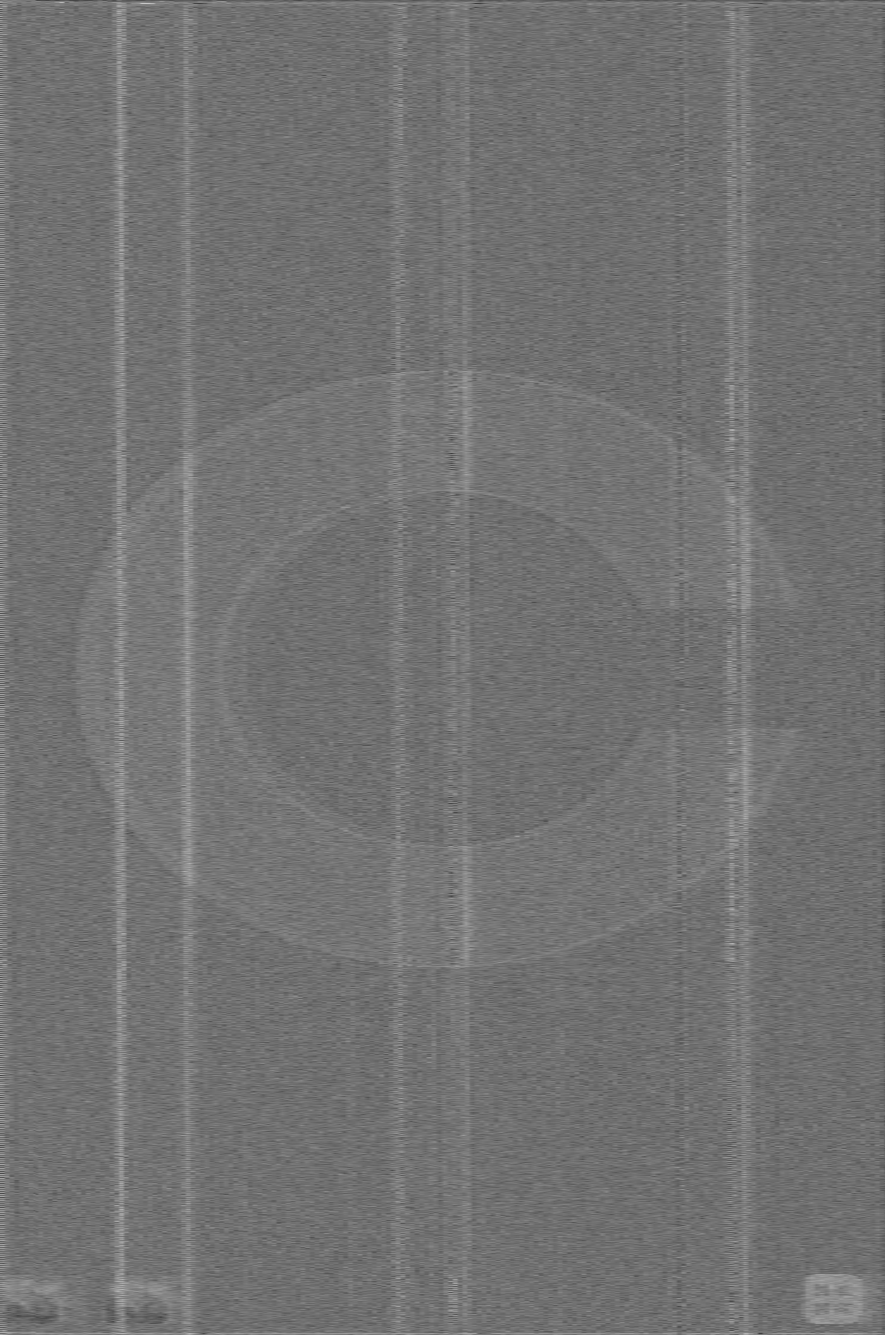}
    \caption{20x}
    \label{fig:emagec10_appendix}
  \end{subfigure}%
  \caption{Images (top row) and their corresponding emages (bottom row) of letter C displayed at 11 different scales. It can be seen that the scales span from uninterpretable to the human eye to easily interpretable.}
  \label{fig:eyedoc_images}
\end{figure*}

In practice, the localization of security code patterns in either time or space dimensions is crucial.
Here we discuss how a simple sliding window technique can tackle both.
When monitoring the target phone in real-time, we can also integrate our recognition model with a simple sliding window operation to identify the key frame(s) that are most likely to contain a text message of the security code.
Specifically, we set the height of the sliding window as the height of each digit, and the width as the total width of 6 digits. The horizontal and vertical strides are equal to the height and width of each digit.

As shown in Figure~\ref{fig:slidingwindow}, the message area is activated much more than the plain area, indicating that our recognition model can be used to identify the most likely frame(s).
Furthermore, within the specific row of the text message, the highest activation responses are concentrated on the security code region.
This suggests that the textual background will not interfere in our security code recognition.
It is also worth noting that, in practice, the attacker could also leverage off-the-shelf language models or visual detection models.
Such models would provide a straightforward way to boost the localization performance.
We also mention that in our experiments, we use a maximum contrast between the background and the text.
Reducing the contrast leads to a less easily readable screen for the human eye, but does not necessarily result in an emage that is more difficult to interpret.
Exploratory experiments confirmed that our choice of background represents a challenging setting, and that, if the attacker is lucky, the contrast between the background and the message on the display of the phone might actually make the attack easier.

\section{Testbed} \label{sec:testbed}
So far, we have introduced the screen gleaning attack and shown its effectiveness in recovering a security code displayed as a push message on the screen of a mobile phone.
The attack was carried out with technology representative of the current state of the art.
However, with time, we expect the quality of the antenna and SDR to improve.
Also, additional training data and algorithmic advances will increase the accuracy of the deep learning classifier.
These advances mean that screen gleaning attacks can be expected to become increasingly dangerous, and future work will be necessary to understand them and develop countermeasures.
To support this future work, we have developed a testbed that enables the systematic test of screen gleaning attacks under incrementally more challenging attacker models.
In this section, we describe the testbed, which has also been released so that can be directly used by the scientific community.
The testbed consists of two parts, first, a definition of a set of images and a set of scales, and, second, a specification of the attacker model, in terms of the model dimensions and the parameterization of the dimensions.
We also report the results of experiments validating the testbed.

\subsection{Testbed Images}
We base our testbed on the eye chart used by eye doctors to test vision acuity~\cite{snellenchart}, and for this reason, we call it the \emph{eye chart testbed}.
Most people are familiar with the experience of a vision test.
The eye chart measures someone's vision by determining the minimum level of detail that the person's eyes can distinguish at a given distance.
Likewise, our testbed uses eye chart letters to determine the minimum level of visual detail that a screen gleaning attack can recover given a particular attack setup.

The testbed is deployed by first specifying an attacker model and creating an attack setup based on that model.
Then, different scales are tested until it is possible to determine at which scale the identity of the letter can no longer be recovered by the attack.
\begin{table}[t]
\normalsize
\newcommand{\tabincell}[2]{\begin{tabular}{@{}#1@{}}#2\end{tabular}}\resizebox{\columnwidth}{!}{
\begin{tabular}{L{.23\linewidth} R{.77\linewidth}}
\toprule
\textbf{Dimension}&\textbf{Description}\\\midrule
Message&The symbol set (e.g., 0-9, a-z) must be defined. If the symbols are not all equiprobable, the prior probability of each symbol must be defined.\\\midrule
\tabincell{l}{Message\\appearance}&Any constraints that will be imposed on the scale of the message or on font types must be defined. Assumptions about the pattern of the background and the brightness of the screen must be defined.\\\midrule
\tabincell{l}{Attack\\hardware}&The antenna and the SDR must be specified. Any assumptions on the position of the antenna must be defined (positions range from touching the phone, to under the table, to across the room).\\\midrule
\tabincell{l}{Device\\profiling}&The conditions on device access must be defined (attacker has access to the device to be attacked, to devices of an identical model, to devices of the same make). The ability of the attacker to cause a certain image to appear on the accessible devices must be defined, along with the amount of time that the attacker can count on having access. After the number and nature of devices at the attackers disposal is defined, the number and length of the sessions that the attacker can record on each device must also be specified.\\\midrule
\tabincell{l}{Computational\\resources}&Define the amount of time and computational resources available for training, and also for the attack itself (i.e., after the model is trained recovering the message from the emage). \\
		\bottomrule
	\end{tabular}
	}
	\caption{Five-dimensional attacker model: Parameter settings to specify when designing an attacker model for testing with the testbed.}
	\label{tbl:model_design}
\end{table}

The testbed defines 11 different scales.
For the largest scale (20$\times$), the size of the letter is the maximum size that can be fit on the screen, with still leaving 10\% of the letter width as margins on the side.
For the smallest scale, the letter appears with a width of 1/20 of the largest scale.
The relative sizes of the testbed scales are illustrated on the top line of Figure~\ref{fig:eyedoc_images}, using the letter C as an example.
The font is the Sloan font used for eye charts.
We used the Creative Commons licensed version, which is available on Github.\footnote{\url{https://github.com/denispelli/Eye-Chart-Fonts/blob/master/README.md}}
The full letter set in the testbed release is C, D, E, F, L, N, O, P, T, Z.
The full set of letters is tested as each scale.

The letters in an eye chart are chosen so that all the letters in the set are equally easy to read.
This ensures that for each scale, the ability of the person to read the letters is related to the scale, and not to the specific letters.
By choosing to use eye chart letters, we extend this property to our test set.
Different eye charts use different fonts and different letter sets.
We choose our testbed based on the fact that this set is currently in widespread use.

It is natural to wonder why we use the limited set of characters used in an eye chart instead of using a larger set of alphanumeric characters.
The answer is that the testbed is designed to detect the ability of an attack to discriminate and recover visual detail.
Using eye-chart characters means that the results of the testbed reflect the discernability and interpretability of other forms of visual information as well, for example, symbols or images displayed on the phone screen, and not just text.

Figure~\ref{fig:eyedoc_images} depicts emages that were captured with the setup described in Section~\ref{sec:attack_setup}.
It can be seen that they move from being uninterpretable to the human eye on the left to interpretable on the right.
This property of the testbed has the goal of ensuring that the testbed can measure interpretability with other attack setups.
We are especially interested in supporting the investigation of attack setups where the signal might be very weak, for example, as the antenna is moved further from the phone.
For a very weak signal, the larger letters will become uninterpretable to the human eye.
This will allow researchers to quantify the effectiveness of a machine-learning attack under the conditions of a weak signal.
If researchers adopt the same standard testbed, the measurements made can be more easily compared in a fair manner.

Again, it is important to note that although our testbed consists of letters, it does not specifically assess the ability of an attack to recover written text consisting of letters.
Instead, it assesses the ability of the attack to recover a message that has a certain level of visual detail.
Just like the eye chart tests general visual acuity, and not just reading, our testbed tests the acuity of a particular attack to recover information in the visual form displayed on the phone screen, and not just letters.

\subsection{Parameterization of the Attacker Model}
Here, we describe the parameterized attacker model.
It contains five dimensions, message, message appearance, attack hardware, device profiling, and computational resources.
Each of these dimensions has several parameters.
In order to have a fully specified attack mode, specifications must be made for each of the parameters.
The parameters can be considered to correspond to the values of design decisions.
The five dimensional model along with the parameters for each dimension are described in Table~\ref{tbl:model_design}.
Note that in the security code attack we present in Section~\ref{sec:attacker_model}, we use the same five dimensions in the attacker model (Table~\ref{tbl:our_attacker_model}).

This parameterized attacker model forms the basis for the attack setup. It has two purposes. First, it ensures that when the testbed is being applied, the attacker model that is being assumed is fully described, i.e., no detail is left out.
Second, it allows researchers to systematically make the attack stronger.
The attack strength can be increased by increasing the values of any or all the parameters.
In this way, the attacker model guides researchers in discovering increasingly strong attacks.
The dimensions of the attacker model can be also used to guide the development of countermeasures.

\subsection{Validating the Eye Chart Testbed}
In this section, we validate the eye chart testbed with the demonstration of an attack.
The attack uses the same Attack hardware and Computational resources as the Security Code attack demonstrated in Section~\ref{sec:attacker_model}.
The Message and the Message appearance are derived from the eye chart testbed.
The Device Profiling is also the same, and the specifics of data collection are explained in the next section.

\subsubsection{Data Collection}

\begin{figure}
	\centering

	\includegraphics[width=0.8\linewidth]{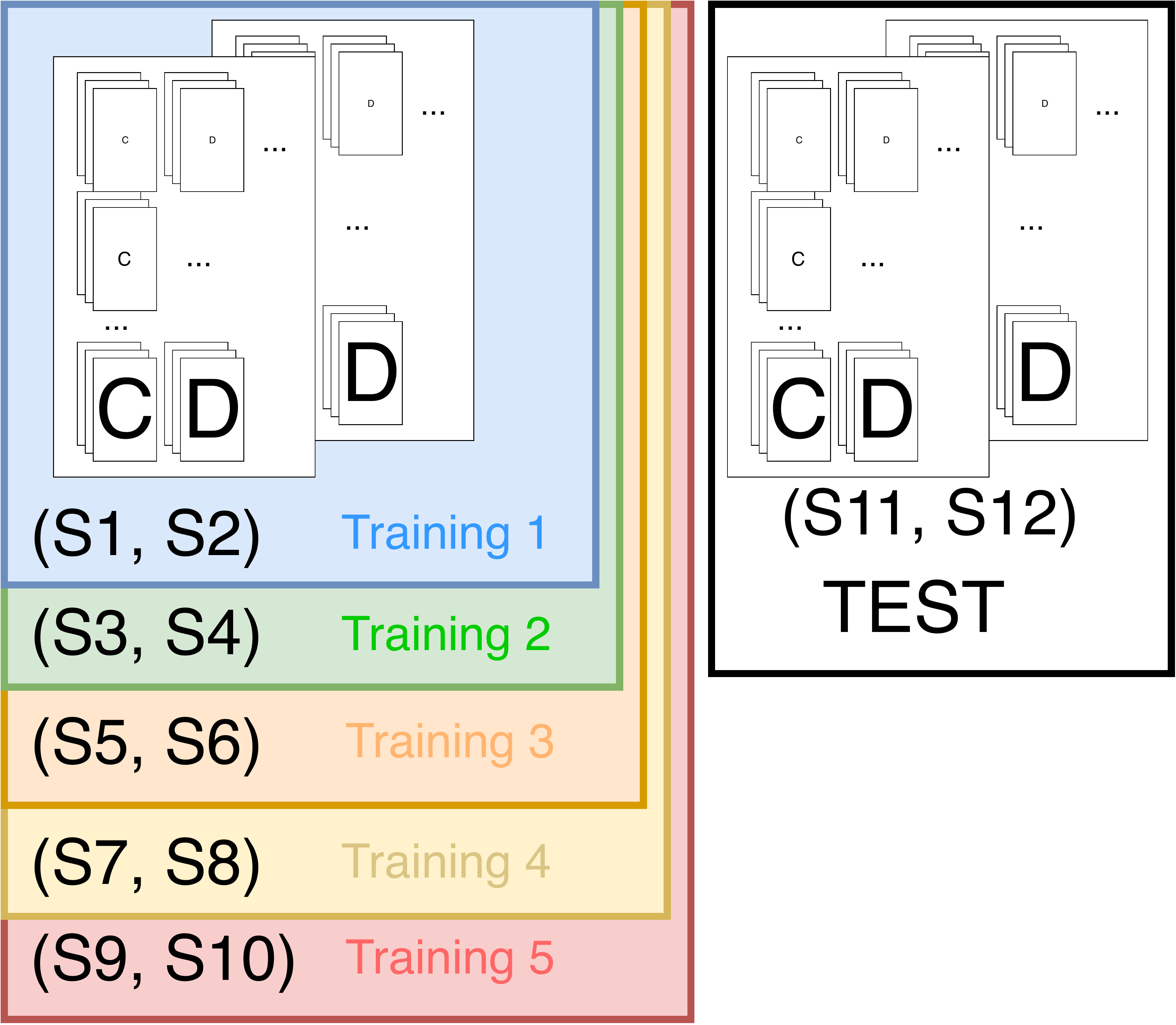}
	\caption{Train/test splits specified in the inter-session case of our eye chart letter classification task, where the training set is gradually enlarged by including more sessions, and the test set is fixed with two sessions.\label{fig:eyedoctordata}}
\end{figure}

\begin{figure}[t]
	\centering

	\includegraphics[width=0.8\columnwidth]{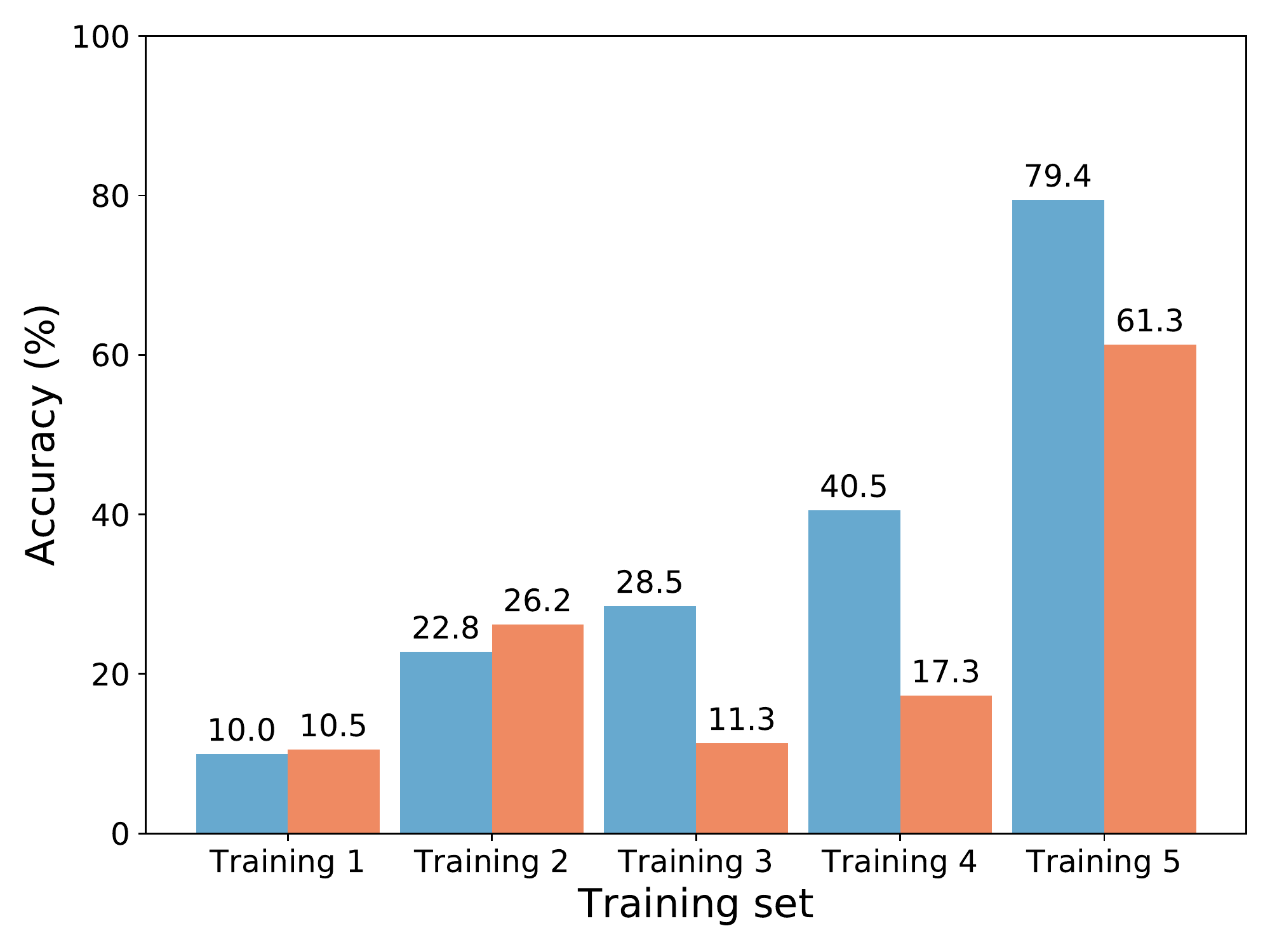}
		\caption{Inter-session accuracy in our eye chart letter classification task. The two bars for each training set represent two different test sessions.\label{fig:corss_session_acc}}
\end{figure}

\begin{figure}[t]
	\centering
	\includegraphics[width=0.8\columnwidth]{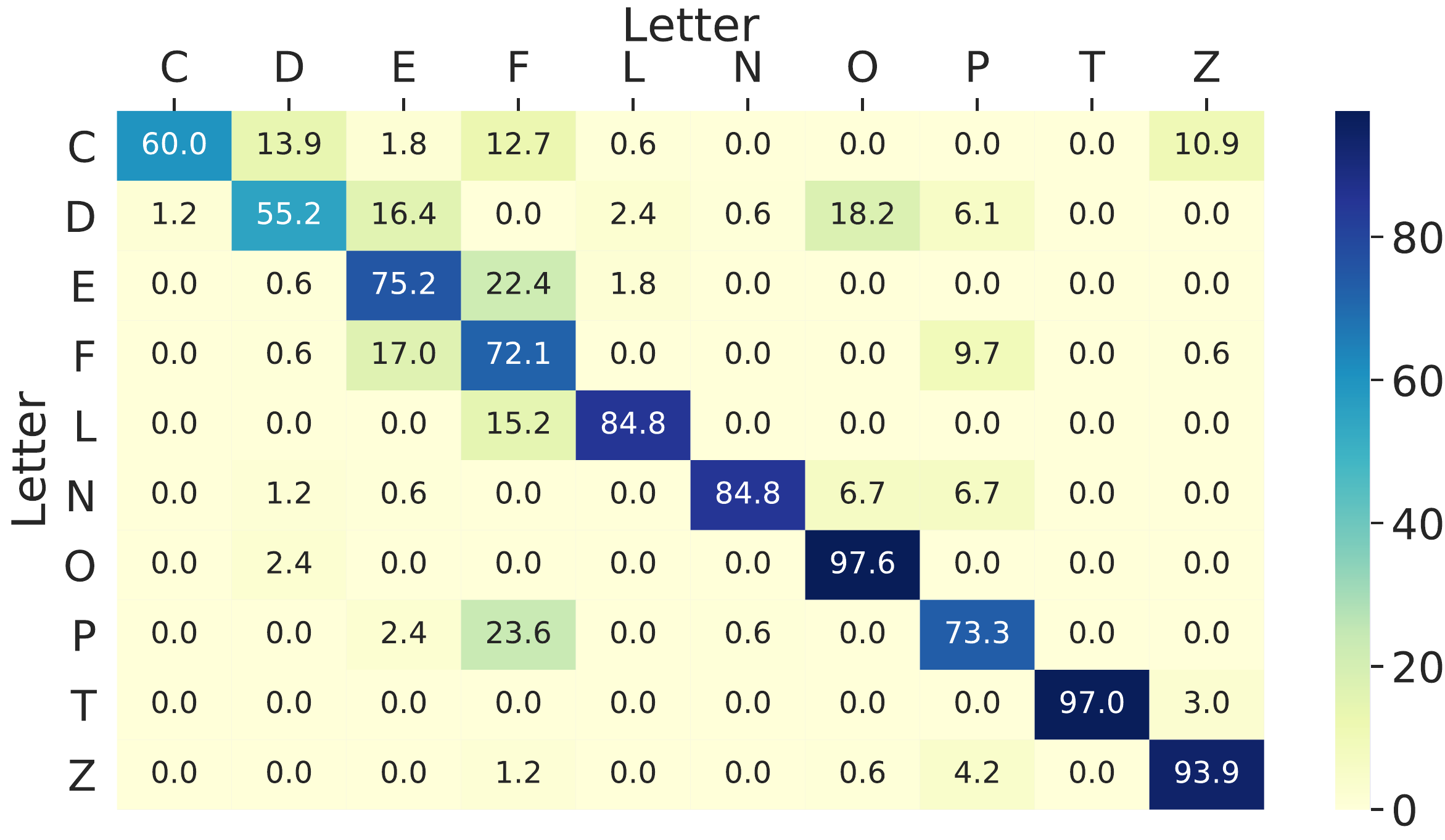}
		\caption{Confusion matrix of the classification in our eye chart letter task.\label{fig:ct_ct}}
\end{figure}

We collect a total of 12 sessions, among which two sessions (sessions 1 and 2) have 50 samples for each of 11 classes of letters at each of 10 scales, and the rest 10 sessions have 15 per class per scale.
For inter-session evaluation, we use sessions 11 and 12 as testing sets for all the experiments.
Session 1 plus 2 are used as the initial training set, and are gradually enlarged by including two more sessions each time, resulting in five different training sets with increased size, denoted as Training 1, 2, 3, 4 and 5, as illustrated in Figure~\ref{fig:eyedoctordata}.
Training 5 has the most data with 24200 samples.
\subsubsection{Experiments}
Similar to the security code attack, we use the following partitions: 80\% training, 10\% validation and 10\% testing.
We use the ResNet-18 model~\cite{he2016deep} as our classification model, and we train on five training sets individually until convergence.
Figure~\ref{fig:corss_session_acc} shows that including more training sessions generally lead to performance improvement in the inter-session case.
For the second session, we notice an accuracy drop when including more data from Training 2 to Training 3, which can also be explained by the fact that the data quality of different sessions of data could impact the performance.

\begin{figure}[t]
	\centering

	\includegraphics[width=0.8\columnwidth]{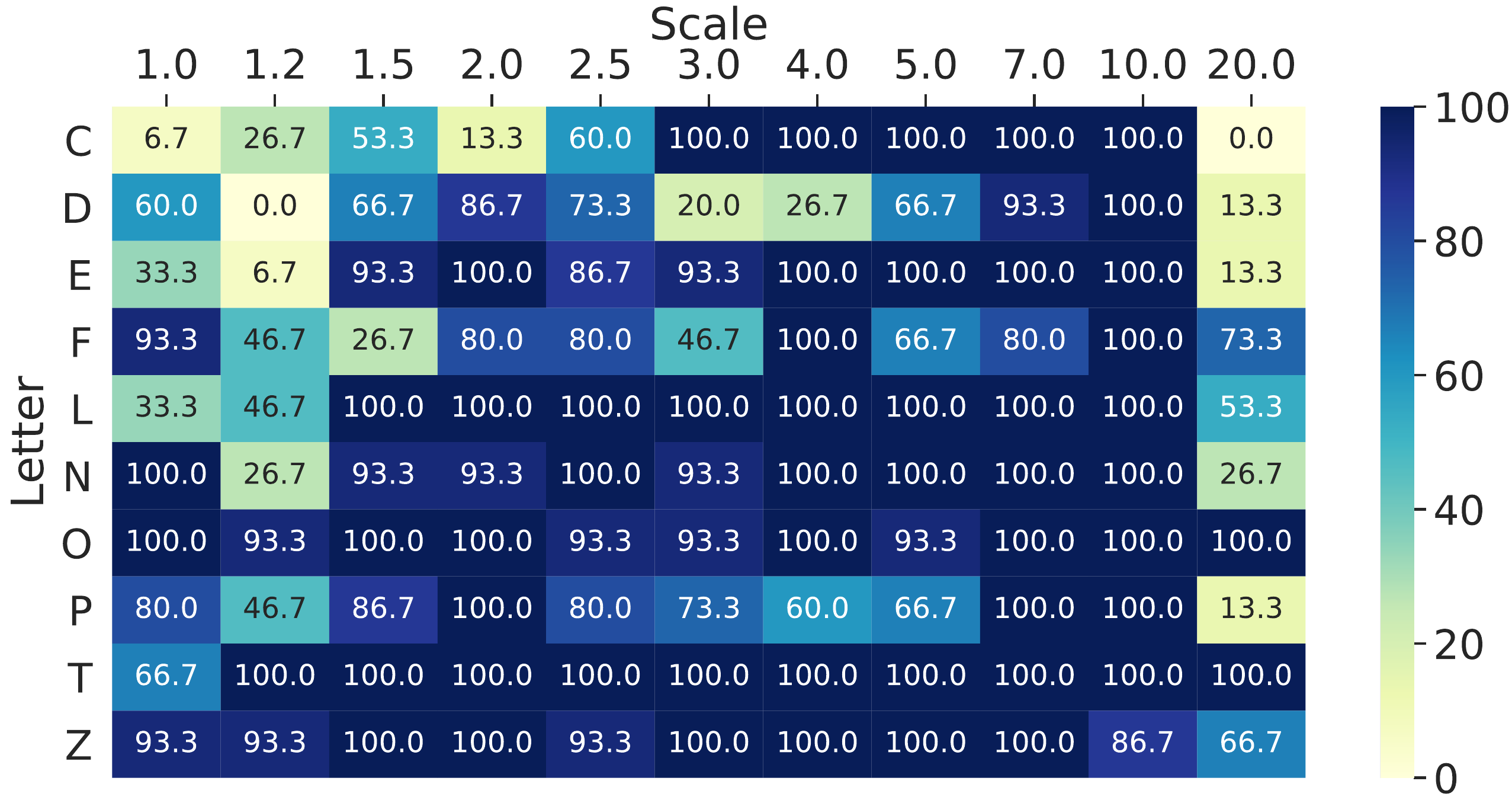}
	\caption{Classification results of our eye chart letter task with respect to different classes and scales.\label{fig:ct_scale}}
\end{figure}

Figure~\ref{fig:ct_ct} shows the confusion matrix of the classification accuracy with respect to different classes.
We can observe that accuracy differs for different letters.
Table~\ref{tbl:scale} shows the results at 11 different scales.
We could observe that the accuracy of the letters at moderate scales (e.g, 7, 8 and 9) is comparatively higher than the others.
Without surprise, the smallest scale has the lowest accuracy.
However, what we found also interesting is that accuracy with respect to scale 1 is also low.
We suspect that it is because of the receptive field of the model we chose.
More detailed results per class per scale can be found in Figure~\ref{fig:ct_scale}.

\begin{table*}[t]
	\centering
	\begin{tabular}{c|ccccccccccc}
		\toprule

		\textbf{Scale}& 1  &1.2     &1.5    &2    &2.5  &3&4   &5   &7 &10     &20 \\
		\midrule
		\textbf{Acc. (\%)} & 66.7 & 48.7 & 82.0 & 87.3 & 86.7 & 82.0 & 88.7 & 89.3 & 97.3 & 98.7 & 46.0\\
		\bottomrule
	\end{tabular}

    \caption{Accuracy with respect to 11 different scales in our eye chart letter classification task.}
	\label{tbl:scale}
\end{table*}

\section{Countermeasures}
\label{sec:discussion_countermeasures}
In our setup, the target device has no extra protection beyond the common design features of commercial devices.
As a step towards improving the protection of the device, we discuss possible countermeasures that could possibly mitigate the danger of a potential screen gleaning attack.

\subsection{Hardware-Based Approaches}
Screen gleaning attacks would be made difficult by using a shielding technique.
Shielding a cable consists of wrapping the center core of the cable that transmits an electric signal by a common conductive layer.
The shield acts as a Faraday cage inside the cable, blocking electromagnetic waves. The resulting electromagnetic leakage is lowered, decreasing the SNR of the signal.
Several standard cables (e.g., coaxial cable, twisted pair cable) are shielded to reduce its electromagnetic perturbations and emanations.
However, this technique comes at an extra cost and increases the cable dimension.
For this reason, flexible flat cables inside small electronics with a display often lack a protective shield, and it is not trivial to add one.

A metallic protective case would also act as a shield for electromagnetic radiation, preventing attacks that measure the signal emitted from the back of the phone, but every telecommunication signal would also be perturbed.

\subsection{Communication-Based Approaches}
Another countermeasure against screen gleaning, similar to the method used for pay-TV, could be to encrypt the signal between the graphical unit and the screen.
The core idea is to share a cryptographic key between the two entities and encode the video stream using a cipher.
As a result, the leaked information by the transmitted signal will become more difficult to interpret by the attacker, who does not have the key.
This solution comes at a cost.
Although some stream ciphers could meet requirements for throughput and latency, both the screen and the graphical unit would need extra logic for encryption and decryption of the cipher and implement a key establishment protocol to create a shared key when paired together.
Moreover, this countermeasure would be ineffective against an attack targeting the screen itself during the rendering (although this is a different attack, see~\cite{genkin2019synesthesia}).

\subsection{Graphics-Based Approaches}
M. G. Kuhn in~\cite{kuhn1998soft} introduces a cheap and efficient countermeasure against electromagnetic TEMPEST that consists of a special font where the transmitted signal has been filtered to reduce the strength of the top peaks of its Fourier transform.
The resulting font appears visually quite blurry for a high-resolution representation rendered on the screen but makes the side-channel silent.

Another method that can be used as a countermeasure is obfuscation.
This obfuscation can either be introduced into the background of the image using confusing patterns and colors behind the text or by using a font with visually difficult to differentiate letters.
However, obfuscation is often ineffective against distinguishing methods based on machine learning and may introduce difficulties for humans to read the original image from the screen.

\section{From Text to Image}
\label{sec:text2images}
Here we return to the discussion of different formulations of the screen gleaning problem.
As we stated earlier, in the discrimination scenario, the attacker knows a finite set of messages that are possible and attempts to determine which one actually occurred on the phone screen.
The security code recovery attack belongs to the discrimination scenario.

As the work on screen gleaning moves forward, it is interesting to look at problems beyond recovering messages built from symbol sets, such as security codes and written words, but also at images.
Screen gleaning of images can be addressed within the reconstruction scenario, mentioned above.
In this scenario, the attacker has no prior knowledge of the screen contents and attempts to reconstruct the screen exactly as it appears to the human eye.
The following is an example of the reconstruction scenario: If the screen was displaying a photo of a person, the goal of the attack would be to recover that photo completely.
Complete recovery requires that the features of the person in the photo are clear, as needed for a human viewer to identify the person, but also that the recovered photo looks exactly like the original one including details of the background and the lighting and coloring of the photo.

Screen gleaning of images can also be addressed within a more general classification scenario than the discrimination scenario.
The discrimination scenario is a type of classification scenario in which the attacker has access to information about the complete set of possible messages.
There exists another classification scenario, which we call the \emph{generalization scenario}, in which the attacker only has some information about the possible content of the screen.
Pornography detection is an example of a problem that needs to be addressed in the generalization scenario.
We discuss it in more detail here because of its societal relevance, cf. the issue of people looking at porn on their devices on an airplane~\cite{nyt2011, simpleflying2020}.

For pornography detection, the attack goal is to determine whether or not a phone display pornography without a direct line of sight to the phone.
Here, we assume, it is not possible for the attacker to have complete information in advance about all possible images displayed on the phone.
Even if it is possible to access a complete database of all pornographic images, it is not possible to know which non-pornographic images will be displayed.
To mount a screen gleaning attack in this case, we must collect representative training data of the different types of phones we expect, similarly to the discrimination case, and different levels of favorability for antenna positioning.
We also, however, must collect representative data of all the different types of pornographic and non-pornographic images that could be relevant to the problem.
The data collection task is clearly not trivial.
However, this type of scenario is clearly important, so we recommend that future work on screen gleaning focuses not only on discrimination scenarios (as with the security codes) but also on more general classification scenarios (as with pornography detection).

We have based our proposed testbed on a test used for visual acuity, and not specifically for reading.
We have made sure that our testbed is not limited to letters and numbers, since we hope that, moving forward, the testbed will be useful for testing screen gleaning in classification scenarios involving generalization and reconstruction.
However, assessing the true capacity of our testbed will require validation tests in addition to those carried out here.

\section{Conclusion and Outlook} \label{sec:conclusion}
In this paper, we have introduced screen gleaning, a new TEMPEST attack that uses an antenna and software-defined radio (SDR) to capture an electromagnetic side channel, i.e., emanations leaking from a mobile phone.
We demonstrate the effectiveness of the new attack on three different phones with an example of the recovery of a security code sent in a text message by using machine learning techniques, as the message is not comprehensible to the human eye.

In addition, we propose a testbed that provides a standard setup in which screen gleaning can be tested further with different attacker models.
Finally, we provide ideas for possible countermeasures for the screen gleaning threat and discuss their potential.

Future work will involve testing increasingly sophisticated attacker models that can be built by extending the five dimensions of the parameterized model that we propose as part of our testbed framework.
As already mentioned, such an extension will involve moving to more sophisticated attack hardware, as hardware continues to develop.
We have already identified special electromagnetic near-field scanners~\cite{EMSCAN}, which are basically arrays of loop antennas that allow the attacker to identify the `hot spot' of the device. The attacker is then able to aim the antenna at this particular spot. These near-field scanners also identify all resonating frequencies within a band of 15 kHz to 80 GHz. These frequencies could then be used for the design of antennas that extend the setup such that attacks on greater distance can be performed.

Further, we will consider a wider range of other devices, including
other screens from devices like tablets, laptops and smart displays (such as Google Nest Hub).
For example, the work of Enev et al.~\cite{enev2011televisions} suggests that our conclusion should remain valid for most of the screens, including TV screens.

Finally, we are interested in moving from discrimination scenarios to generalization scenarios, and finally to reconstruction scenarios.
In other words, content that the attack recovers from the phone will become increasingly unpredictable, and increasingly challenging.
The testbed we presented here has the potential to be further developed to also cover the full range of possible scenarios.

\section*{Acknowledgments}
Part of this work was carried out on the Dutch national e-infrastructure with the support of SURF Cooperative.
We thank Peter Dolron and Daniel Sz\'alas-Motesiczky of the TechnoCentrum at Radboud University for their support with the measurement setup.
A special word of appreciation to Frits, Henan, Jan, Maikel, and Mia, who contributed time with their phones, so that we could carry out screen gleaning attacks.

\bibliographystyle{IEEEtranS}
\bibliography{jobname}

\end{document}